\documentclass[prd,aps,superscriptaddress,nofootinbib,preprint,eqsecnum,showkeys,tightenlines]{revtex4-1}
\usepackage{slashed}
\usepackage{epsfig,latexsym,cancel,amssymb,amsmath,verbatim,mathrsfs}
\usepackage{color}
\usepackage{graphicx}

\def\Ld{\Lambda}

\def\f{\frac}
\newcommand{\be}{\begin{equation}}
\newcommand{\ee}{\end{equation}}
\newcommand{\bea}{\begin{eqnarray}}
\newcommand{\eea}{\end{eqnarray}}
\newcommand{\ba}{\begin{array}}
\newcommand{\ea}{\end{array}}

\long\def\symbolfootnote[#1]#2{\begingroup%
\def\thefootnote{\fnsymbol{footnote}}\footnote[#1]{#2}\endgroup}

\newcommand{\beq}{\begin{equation}}
\newcommand{\eeq}{\end{equation}}

\newcommand{\ET}{\mbox{$\not \hspace{-0.10cm} E_T$}}

\definecolor{My_red}{cmyk}{0.00,1.00,1.00,0.40}

\newcommand{\slg}[1]{#1\hspace{-0.45em{/}}}
\newcommand{\slG}[1]{#1\hspace{-0.6em{/}}}

\begin{document}

\title{Signatures of a Flavor Changing $Z'$ Boson in $B_q \to \gamma Z'$}

\author{Shao-Long Chen}
\email[E-mail: ]{chensl@mail.ccnu.edu.cn}
\affiliation{Key Laboratory of Quark and Lepton Physics (MoE) and Institute of Particle Physics, Central China Normal University, Wuhan 430079, China}

\author{Amit Dutta Banik}
\email[E-mail: ]{amitdbanik@mail.ccnu.edu.cn}
\affiliation{Key Laboratory of Quark and Lepton Physics (MoE) and Institute of Particle Physics, Central China Normal University, Wuhan 430079, China}

\author{Zhaofeng Kang}
\email[E-mail: ]{zhaofengkang@gmail.com}
\affiliation{School of Physics, Huazhong University of Science and Technology, Wuhan 430074, China}

\author{Qin Qin}
\email[E-mail: ]{qqin@hust.edu.cn}
\affiliation{School of Physics, Huazhong University of Science and Technology, Wuhan 430074, China}

\author{Yoshihiro Shigekami}
\email[E-mail: ]{sigekami@post.kek.jp}
\affiliation{School of Physics, Huazhong University of Science and Technology, Wuhan 430074, China}
\date{\today}

\begin{abstract}
Rare $B$ meson decays offer an opportunity to probe a light hidden $Z'$ boson. 
In this work we explore a new channel $B_q\to \gamma Z'$ ($q = d, s$) followed by a cascade decay of $Z'$ into an invisible (neutrino or dark matter) or charged lepton pair $\ell^+\ell^-$ ($\ell=e ,\mu)$. 
The study is based on a simplified effective model where the down quark sector has tiny flavor-changing neutral current couplings with $Z'$. 
For the first time, we calculate ${\rm BR}(B_q\to \gamma Z')$ at the leading power of $1/m_b$ and $1/E_\gamma$. 
Confronting with the strong constraints from semi-invisible decays of $B$ meson, we find that the branching ratio for $B_d \to {\rm invisible}+ \gamma$ can be larger than its Standard Model prediction, leaving a large room for new physics, in particular for light dark matter. 
Additionally, the branching ratio for $B_d \to e^+e^- \gamma$ can also be sizable when the corresponding flavor violating $Z'$ coupling to quarks is of the axial-vector type. 
On the other hand, the predicted branching ratios of $B_d \to \mu^+\mu^- \gamma$ and $B_s \to \ell^+ \ell^- \gamma$ are severely constrained  by the experimental measurements.

\end{abstract}


\maketitle

\section{Introduction}

At the energy frontier, the large hadron collider (LHC) has successfully reached the TeV scale at the parton level.
However, there are still null results for the hunting of new fundamental particles. The TeV-energy frontier is probably the brightest beacon for new physics searching, but on the contrary, a faint lighthouse may still be set up by high-intensity low-scale experiments. 
This motivates the community a lot to search for light and very weakly interacting dark particles, and we refer to the report~\cite{SHiP} for a review on the ongoing and proposed experiments.

The light $Z'$ boson, a new gauge boson associated with an additional certain gauge symmetry, is a benchmark candidate. 
Its mass can be varied within a large range and has been extensively studied in the literature.
A low-scale $Z'$ is well motivated, for instance, by the strong exclusion to dark matter above the GeV scale; a light $Z'$ below the GeV scale could provide a portal to light dark matter~\cite{Kang:2010mh}. 
In addition, a light $Z'$ can mediate large self-interactions between dark matter thus solving the cosmological small scale crisis~\cite{Weinberg:2013aya}. 
A $Z'$ boson with mass of $\mathcal{O}(10 \mathchar`- 100)$ MeV is well motivated to alleviate the long-standing tension of muon anomalous magnetic moment $(g-2)_{\mu}$~\cite{g-21,g-22,g-23,g-24,g-25,g-26,g-27,Altmannshofer:2016brv,Kang:2019vng} and a GeV-scale $Z'$ is proposed to account for the anomalies in $B$ rare decays~\cite{Sala:2017ihs,Ghosh:2017ber,Altmannshofer:2017bsz,Darme:2020hpo}. 

Generally speaking, a $Z^{\prime}$ boson could induce flavor-changing neutral currents (FCNCs). The FCNC processes induced by a heavy $Z'$ around the weak scale have been exclusively studied~\cite{Langacker:2000ju,Blake:2016olu}. 
In this work, we study the FCNC processes induced by a light (hidden) $Z^{\prime}$ faintly coupling to the down quarks with tree-level FCNC. Whereas the possible radiative contribution from the FCNCs in the up quark sector is negligible, a situation naturally arising in a class of gauged flavorful model.

It is well known that in the Standard Model (SM) there are no FCNCs at tree-level, while the FCNC transitions occurring through a $W$-loop, within the framework of Cabibbo-Kobayashi-Maskawa (CKM) charged current mixing, are strongly suppressed by the Glashow-Iliopoulos-Maiani (GIM) mechanism~\cite{GIM}. 
Therefore, searching for rare decays provides a sensitive probe to new physics beyond the SM which may just slightly violate the flavor structure of the SM. 
In particular, if the flavor changing effects are related to the down quark sector, the light $Z'$ may leave its fingerprints in the rare decays of $B$ mesons at the $B$-factory, the SuperKEKB which has started the phase 3 since March, 2019. 
It has been investigated, since many years ago, for a light dark photon~\cite{Pospelov:2008zw,Davoudiasl:2012ag}. 
Despite of the absence of tree-level FCNC, rare decays $B \to K Z'(\to \ell^+ \ell^-,\,\nu\bar\nu)$ ($\ell=e, \mu$) can be induced via the $W$-loop. 
We noticed that possible signatures of a light dark matter with flavor changing coupling to bottom quark have been studied by Refs.~\cite{Badin:2010uh,Fernandez:2015klv}. 
A study of $B_c$ decaying into light invisible particle was done recently by Li {\it et al.} \cite{Li:2018hgu}. 
Dark matter searches with low mass scalar mediator in $B$ decays was carried out in Refs.~\cite{Schmidt-Hoberg:2013hba,Filimonova:2019tuy}.

In this work, we focus on a new decay channel $B_q \to \gamma Z' (q=d ,s)$ 
with an on-shell $Z'$ followed by prompt decays $Z' \to \ET$ or $\ell^+ \ell^-$, which offers a new way to probe a light $Z'$. However, the resulting signatures are not completely new and some related studies have already been done before. For the radiative dilepton signatures $B \to\gamma \ell^+ \ell^-$, aside from the studies within the SM~\cite{Aliev:1996ud,Geng:2000fs,Kruger:2002gf,Melikhov:2004mk,Kozachuk:2017mdk,Kozachuk:2018dqc}, new physics effects beyond the SM are also studied~\cite{Xiong:2001up,Aliev:2001dj,Heng:2008rc,Dettori:2016zff,Banerjee:2019cze,Abbas:2018xdu}. 
The signature of single photon plus missing energy is also studied and may be of more interest since it is related to light dark matter~\cite{Badin:2010uh}. All of these studies are based on heavy new physics where the mediators accounting for the FCNC decays of $B$ are integrated out; on the contrary, our study is based on a light $Z'$ boson with mass of MeV-GeV range.

Experimentally it is challenging to detect the radiative $B$ decays, because the additional photon is difficult to detect, and what is worse, it softens the charged leptons thus hampering their reconstruction. 
So far, the BaBar collaboration searched for $B_d \to \gamma + \ET/ \ell^+ \ell^-$~\cite{Aubert:2007up,Lees:2012wv}~\footnote{These searches are with respect to signatures from the effective operators, which are different from our scenario with a light resonance.}, 
 and recently the Belle collaboration reported updated results for $B_d \to \gamma + \ET$ \cite{Ku:2020aix}. 
The resulting upper bounds are well above the SM predictions. 
Whereas the Belle II physics book does not discuss their prospects~\cite{Kou:2018nap}. At the LHCb, to our knowledge, only one note~\cite{Bonivento:2010nza} studied the $\mu^+\mu^-$ mode, finding that the LHCb experiment will be able to put an upper limit on BR($B_s \to \mu^+ \mu^- \gamma)\lesssim 6.0\times 10^{-9}$ with 2~$\rm fb^{-1}$ integrated luminosity. On the other hand, the channel $B_q \to \gamma Z'$ is accompanied by decays $B \to P/V Z'$ with $P=\pi, K$ and $V=\rho, K^*$. The $P$ mode can put bounds on the vector FCNC couplings $\bar{q} \gamma^\mu b Z'_\mu$, while the $V$ mode puts bounds on the axial-vector couplings $\bar{q} \gamma^\mu\gamma^5 b Z'_\mu$. Considering such a situation, we merely explore the remaining room for $B_q \to \gamma Z'$, conservatively requiring that $B_q \to \gamma Z'(\to \bar\nu\nu/ \ell^+ \ell^-$) are not buried beneath the corresponding signatures of the SM. 
We systematically analyze the available experimental results from BaBar, Belle and LHCb, and find that they put strong limits on the allowed branching ratio of $B_q \to \gamma Z'$, except when $Z'$ mass is close to the $B$ meson mass. 
However, there is still a relatively wide parameter space for the missing energy mode and the $e^+ e^-$ mode for $B_d$. 
Improving sensitivities to these signatures may be urgent. See a recent attempt at the LHCb~\cite{Dettori:2016zff}.

The paper is organized as the following: 
In Section~\ref{sec:setup} we present the (effective) models that describe FCNC in the down quark sector involving $b$. 
In the next section we calculate the branching ratio of $B_q \to \gamma Z'$. 
In Section~\ref{sec:analysis} we analyze the SM backgrounds of the resulting signatures and their prospects. Section~\ref{sec:conclusion} is devoted to the conclusion.

\section{FCNC transitions $b \to d/s$ induced by a light $Z'$}
\label{sec:setup}

FCNC transitions can occur either at tree-level or at loop level, and the latter can transfer FCNCs in the up quark sector to the down quark sector with the help of a $W$-loop, which is of special interest considering that the up quark sector allows a relatively larger FCNC. 
However, in this paper we assume that FCNCs induced by a light $Z'$ are presented in the down quark sector at tree-level, to investigate the discovery potential of such a flavor changing light $Z'$ from $B$-meson rare decays.

Let us start with the effective Lagrangian for FCNCs $b \to d/s$ due to a light $Z'$ at tree-level:
\begin{align}
-{\cal L}_{Z'} = \f{1}{2} m_{Z'}^2 Z'_\mu {Z'}^{\mu} + \bar{q} \gamma^\mu \left[(g_L)_{qb} P_L +(g_R)_{qb} P_R\right] b Z'_\mu+h.c.,
\label{eq:LagZ'}
\end{align}
with $q=d, s$. The chiral couplings $(g_{L/R})_{qb}$ are free, but are supposed to be proportional to the gauge coupling associated with the horizontal gauge symmetry $G_f$ such as $U(1)_X$ that presents the massive gauge boson $Z'$. 
Moreover, they involve the elements of the unitary matrix that diagonalizes the down quark mass matrix when we consider flavor dependent $Z'$ couplings and appropriate procedure to accommodate the CKM matrix, as mentioned in Ref.~\cite{Kang:2019vng}. 
$Z'$ is supposed to further couple to the SM fermions such as leptons $\bar{\ell} \gamma^\mu \ell Z'_\mu $ with $\ell = e, \mu, \nu$, and even to a light dark matter $X$. 
However, we do not incorporate the possible couplings to $\bar{q} q$ and $\bar{\tau} \tau$ because the resulting signatures have a poor sensitivity. 
These couplings lead to decays of $Z'$ with the concrete branching ratios depending on the UV models. 
In our analysis, however, for a given channel we will take it to be 100\%, which helps enhance the signature rate. The mass of $Z'$ is a free parameter, and for a very low scale $m_{Z'}$, itself can be a stable particle, at least at the scale of detectors, or the lifetime is just around this scale thus leading to displaced vertex signatures, which are beyond the scope of this work.

Note that in our discussions the possible radiative contributions to $b \to d/s$ from the up quark sector FCNCs are assumed to be highly suppressed. 
Naturally it is true as long as the mass matrix for up quarks is the most generic one allowed by $U(1)_X$; more concretely, there are no additional family mixings from the breaking of $U(1)_X$. 
Then, in the up quark sector, the $Z'$ current and mass matrix can be diagonalized simultaneously.

We would like to comment on the possible UV completions of the effective model. 
A benchmark model is inspired by the muon $g-2$ anomaly~\cite{Kang:2019vng}. 
We briefly describe the model and the FCNC therein. 
It is based on the conventional local $U(1)_{B-L}$, but only the second and third families of fermions are charged under it so as to evade the stringent constraints involving electron. 
Unlike its analogy, the gauged $L_\mu-L_\tau$ model~\cite{LmuLtau1,LmuLtau2} where quarks do not couple to $Z'$ at all~\footnote{Extensions to the gauged $L_\mu-L_\tau$ model giving FCNC are also considered in some studies~\cite{FVLmuLtau1,FVLmuLtau2,FVLmuLtau3,FVLmuLtau4,FVLmuLtau5}.}, this model can give rise to FCNC currents of $Z'$ in the down quark sector, depending on the way to realize the CKM matrix. 
In the original way, in order to account for muon $g-2$ consistent with the constraints on the down quark FCNCs, the mixings between the first and latter two generations are forced to come from the up quark sector, and then the resulting mass matrices taking the form of
\begin{eqnarray}\label{mass:singlet}
m_u^0 = \f{v_h}{\sqrt{2}}
\begin{pmatrix}
Y^u_{11} & y^u_{12}v_f/\Ld & y^u_{13}v_f/\Ld \\
y^u_{21}v_f/\Ld & Y^u_{22} & Y^u_{23} \\
y^u_{31} v_f/\Ld &Y^u_{32} & Y^u_{33} \end{pmatrix}, ~~~ m_d^0 = \f{v_h}{\sqrt{2}}
\begin{pmatrix} Y^d_{11} & 0 & 0 \\
0 & Y^d_{22} & Y^d_{23} \\
0 &Y^d_{32} & Y^d_{33} \end{pmatrix}, \label{eq:quarkmass}
\end{eqnarray}
where $v_f$ denotes for the spontaneously breaking of $U(1)_X$ by a singlet flavon, whose effect is mediated to the up quark sector via vector-like quarks with quantum numbers similar to the up quarks. 
Then, as stated before, $m_d^0$ and $\bar d \gamma^\mu d Z'_\mu$ can be diagonalized simultaneously. 
However for our purpose, one can modify the above setup by also mediating $U(1)_X$ breaking to the down quark sector, but very slightly; alternatively, we just give up the attempt to explain muon $g-2$ and realize CKM matrix based on Eq.~(\ref{eq:quarkmass}) with $u$ and $d$ exchanged.

\section{The rare decays $B_q \to \gamma Z'$}
\label{sec:calc}

In this section we present the details of the calculation of $B_q \to \gamma Z^{\prime}$. Firstly 
we would like to review the decay $B_q \to K Z'$, which is able to impose strong constraints 
by using the Belle data.

\subsection{Review on $B_q \to K Z'$}
\label{sec:reviewBtoKZp}

From the Lagrangian \eqref{eq:LagZ'}, flavor violating $B$ meson decays happen at tree-level. Its amplitude, for example $B \to K Z'$, is given by
\begin{align}
\mathcal{M} = - i \frac{1}{2} \epsilon^{\prime*}_{\mu} (q) \left[ g_V^{(s)} \langle K | \bar{s} \gamma^{\mu} b | B \rangle - g_A^{(s)} \langle K | \bar{s} \gamma^{\mu} \gamma_5 b | B \rangle \right],
\end{align}
where $\epsilon'_{\mu} (q)$ is the polarization vector of $Z'$ and $g_{V, A}^{(s)} \equiv (g_L)_{sb} \pm (g_R)_{sb}$ are the flavor violating vector and axial-vector couplings, respectively. The hadronic matrix element is defined as~\cite{Gubernari:2018wyi}
\begin{align}
\langle K(p_K) | \bar{s} \gamma^{\mu} b | B(p_B) \rangle = \left[ ( p_B + p_K )^{\mu} - \frac{M_B^2 - M_K^2}{q^2} q^{\mu} \right] f_+ (q^2) + \frac{M_B^2 - M_K^2}{q^2} q^{\mu} f_0 (q^2),
\end{align}
and $\langle K | \bar{s} \gamma^{\mu} \gamma_5 b | B \rangle = 0$. Note that the form factors $f_+ (q^2)$ and $f_0 (q^2)$ must satisfy the relation $f_+ (0) = f_0 (0)$ in order to remove the singularity at $q^2 = 0$. As a result, we obtain the decay width as 
\begin{align}
\Gamma (B \to K Z') = \frac{| g_V^{(s)} |^2}{64 \pi} \frac{\lambda(M_B^2, M_K^2, m_{Z'}^2)^{3/2}}{m_{Z'}^2 M_B^3} | f_+ (m_{Z'}^2) |^2,
\label{eq:decBtoKZp}
\end{align}
where $\lambda(x, y, z) = x^2 + y^2 + z^2 - 2 x y - 2 y z - 2 z x$. It is seen that this decay width merely depends on the vector coupling $g_V^{(s)}$ but not on the axial-vector coupling $g_A^{(s)}$. In the light $Z'$ limit, $\lambda(M_B^2, M_K^2, m_{Z'}^2) \approx M_B^4 \left( 1 - \frac{M_K^2}{M_B^2} \right)^2$, and therefore, Eq.~(\ref{eq:decBtoKZp}) can be simplified to be
\begin{align}\label{BtoK}
\Gamma (B \to K Z') \simeq \frac{| g_V^{(s)} |^2}{64 \pi} \frac{M_B^3}{m_{Z'}^2} \left( 1 - \frac{M_K^2}{M_B^2} \right)^3 | f_+ (m_{Z'}^2) |^2.
\end{align}
It holds for both the neutral and charged kaons, $K^0$ and $K^+$. 
The size of $g_V^{(s)}$ is severely constrained by the current experimental limit: $|g_V^{(s)}| \simeq \mathcal{O} (10^{-10})$ when $m_{Z'} = \mathcal{O}(10)$ MeV \cite{Kang:2019vng}.

Similarly, we calculate the decay width of $B$ into the vector mesons like $B \to K^* Z'$~\cite{Oh:2009fm}
\begin{align}
\Gamma ({B} \to K^* Z') = \frac{\lambda(M_B^2, M_{K^*}^2, m_{Z'}^2)^{1/2}}{16 \pi M_B^3} \left( |H_0|^2 + |H_+|^2 + |H_-|^2 \right),
\label{eq:decBtoKaZp}
\end{align}
where the helicity amplitudes $H_0$ and $H_{\pm}$ are
\begin{align}
H_0 &= g_A^{(s)} \left[ - \frac{1}{2} ( M_B + M_{K^*} ) A_1 (m_{Z'}^2) x_{K^*Z'} + \frac{M_{K^*} m_{Z'}}{M_B + M_{K^*}} A_2 (m_{Z'}^2) \left( x_{K^*Z'}^2 - 1 \right) \right], \\
H_{\pm} &= \frac{g_A^{(s)}}{2} ( M_B + M_{K^*} ) A_1 (m_{Z'}^2) \pm g_V^{(s)} \frac{M_{K^*} m_{Z'}}{M_B + M_{K^*}} V (m_{Z'}^2) \sqrt{x_{K^*Z'}^2 - 1}.
\label{BtoK*}
\end{align}
Here, $x_{K^*Z'} \equiv \left( M_B^2 - M_{K^*}^2 - m_{Z'}^2 \right) / \left( 2 M_{K^*} m_{Z'} \right)$, and $A_1 (q^2)$, $A_2 (q^2)$ and $V (q^2)$ are form factors of the $B \to K^*$ transition. The numerical values for these form factors are taken from Ref.~\cite{Ball:2004rg}. In sharp contrast to $\Gamma (B \to K Z')$, $\Gamma ({B} \to K^* Z')$ instead is sensitive to (dominated by) the axial-vector coupling $g_{A}^{(s)}$, and for $m_{Z'} = \mathcal{O}(100)$ MeV the current bound requires $|g_A^{(s)}| \lesssim \mathcal{O} (10^{-9})$. Therefore, it is a good approximation to quote the bounds on $B \to KZ'$ and $B \to K^*Z'$ as the bounds on the vector and axial-vector couplings, respectively.

The decay width formulas~\eqref{eq:decBtoKZp} and \eqref{eq:decBtoKaZp} can be simply extended to the parallel decay modes $B \to \pi Z'$ and $B \to \rho Z'$ respectively.

\subsection{Calculation for $B_q \to \gamma Z'$}
\label{sec:calcBgZp}

In this subsection, we show the details of the calculation for $\bar{B}_q \to \gamma Z'$ at the leading power of $1/m_b$ and $1/E_\gamma$. 
In the decay, the flavor gets changed via the non-standard $b q Z'$ vertex, and the photon can be emitted from either the bottom quark or the light quark in the $\bar{B}_q$ meson. 
The former case contributes only from the sub-leading power and is neglected in our calculation. 
The rest part of the $\bar{B}_q (P) \to \gamma (k) Z' (k')$ decay amplitude can be formulated as 
\begin{align}
\mathcal{M} = \int d^4x e^{ikx}\left( -{ie\over3}\epsilon_{\mu} (k) \right) \left( -{i\over2}\epsilon'_{\nu} (k') \right)
\langle 0 | T\{ \bar{q}(x)\gamma^\mu q(x)\; \bar{q}(0)\gamma^{\nu}(g_V^{(q)} - g_A^{(q)}\gamma_5) b(0) \} | B_q \rangle \; ,
\label{eq:Amp}
\end{align}
where $\epsilon^{(\prime)}$ is the polarization vector for the photon ($Z'$). 
Similar to the factorization formula for $\bar{B}_q \to \gamma\ell\nu$ \cite{Lunghi:2002ju,Bosch:2003fc} and $\bar{B}_q\to \gamma\gamma$ \cite{Bosch:2002bv}, the above amplitude is factorized at the leading order into 
\begin{align}
\mathcal{M} = -{f_{B_q}\over 4} {e\over 6} \epsilon_{\mu} (k) \epsilon'_{\nu} (k')
\int_0^1 \! d \xi \; {\rm Tr} \! \left[ \gamma^\mu {\slashed{l}\over l^2+i\epsilon}\gamma^\nu (g_V^{(q)} - g_A^{(q)}\gamma_5) (\slG{P} + m_b) \gamma_5 \right] \Phi_{B1} (\xi),
\end{align}
with $l$ being the momentum of the internal light quark propagator, and the light quark mass $m_q$ being omitted. 
The function $\Phi_{B1} (\xi)$ describes the distribution of the light-cone momentum fraction, which is the only non-vanishing 
component of the light-cone projector for the $B$ meson appearing
in our calculation at the leading power \cite{Grozin:1996pq,Lange:2003ff}
\begin{align}
\langle 0 | T \{ \bar{q}_\alpha(x) b_\beta(0) \} | \bar{B}_q(P) \rangle 
\ni 
-{if_{B_q}\over 4}[(\slashed{P} + m_b)\gamma_5]_{\beta\alpha} \int_0^1d\xi e^{-i\xi P_+ z_-} \Phi_{B1}(\xi)\; , 
\end{align}
and it satisfies 
\begin{align}\label{eq:Hadm}
\int_0^1 \! d \xi \; \Phi_{B1} (\xi) = 1\; , ~~~ \int_0^1 \! d \xi \; \frac{\Phi_{B1} (\xi)}{\xi} = \frac{M_B}{\lambda_B} \; ,
\end{align}
where $\xi \equiv p_+ / P_+$ with $p$ ($= k - l$) being the momentum of the constituent light quark in the $B$ meson, and $\lambda_B$ is the first inverse moment of the $B$ meson light-cone distribution amplitude. 
The label `+' denotes the component in the light-cone direction opposite to the external photon momentum $k$. 
Below, we can see why only the `+' component contributes from leading power. 
Since $p \sim \Lambda_{\rm QCD}$ and $k \cdot p \sim m_b \Lambda_{\rm QCD}$, we can expand the propagator as
\begin{align}\label{eq:Gammamunu}
\frac{\slg{l}}{l^2} = \frac{\slg{k} - \slg{p}}{- 2 k \cdot p + p^2} = \frac{\slg{k}}{- 2 k \cdot p} \left( 1 + \mathcal{O} ( {\Lambda_{\rm QCD}}/{m_b} ) \right).
\end{align}
The product $k \cdot p$ picks only the `+' component of $p$, {\it i.e.}, $k\cdot p = k\cdot p_+$. 
Substituting \eqref{eq:Hadm} and \eqref{eq:Gammamunu} into \eqref{eq:Amp}, we further write the amplitude as 
\begin{align}
\mathcal{M} &= \frac{e}{24} \frac{f_{B_q}}{2 k \cdot P_+} \epsilon_{\mu} (k) \epsilon'_{\nu} (k') {\rm Tr} \! \Bigl[ \gamma^{\mu} \slg{k} \gamma^{\nu} \left( g_V^{(q)} - g_A^{(q)} \gamma_5 \right) (\slG{P} + m_b) \gamma_5 \Bigr] \int_0^1 \! d \xi \; \frac{\Phi_{B1} (\xi)}{\xi} \nonumber \\[0.5ex]
&= \frac{e}{6} \frac{f_{B_q}}{2 k \cdot P_+} \Bigl\{ i g_V^{(q)} \varepsilon^{\mu \nu \rho \sigma} \epsilon_{\mu} \epsilon'_{\nu} k_{\rho} k'_{\sigma} + g_A^{(q)} \bigl[ (k \cdot \epsilon') (k' \cdot \epsilon) - (k \cdot k') (\epsilon \cdot \epsilon') \bigr] \Bigr\} \frac{M_{B_q}}{\lambda_{B_q}}.
\label{eq:iM}
\end{align}
The squared amplitude with the polarization summed is given by
\begin{align}
\sum_{\rm pol.} | \mathcal{M} |^2 &= \frac{e^2}{36} \frac{f_{B_q}^2}{(2 k \cdot P_+)^2} \frac{M_{B_q}^2}{\lambda_{B_q}^2} \Bigl\{ 2 | g_V^{(q)} |^2 + 2 | g_A^{(q)} |^2 \Bigr\} (k \cdot k')^2.
\label{eq:Msq}
\end{align}

For the kinematics of this process in the rest frame of the $B$ meson, we have
\begin{align}
k \cdot P_+ = \frac{M_{B_q}^2 - m_{Z'}^2}{2} = k \cdot k', ~~ | \boldsymbol{k} | = | \boldsymbol{k'} | = \frac{M_{B_q}^2 - m_{Z'}^2}{2 M_{B_q}} = \frac{M_{B_q}}{2} ( 1 - r_{Z'} ),
\label{eq:kin}
\end{align}
where $k = (E_{\gamma}, \boldsymbol{k})$ and $k' = (E', \boldsymbol{k'})$, and we define $r_{Z'} \equiv m_{Z'}^2 / M_{B_q}^2$. As a result, we obtain the decay width of $B_q \to \gamma Z'$ as
\begin{align}
\Gamma (B_q \to \gamma Z') &= \frac{1}{8 \pi} \sum_{\rm pol.} | \mathcal{M} |^2 \frac{| \boldsymbol{k} |}{M_{B_q}^2} \nonumber \\
&= \frac{\alpha}{32} \frac{M_{B_q}}{9} \frac{f_{B_q}^2}{\lambda_{B_q}^2} \Bigl( | g_V^{(q)} |^2 + | g_A^{(q)} |^2 \Bigr) ( 1 - r_{Z'} )\,.
\label{eq:GamBqtogammaZp}
\end{align}
For the inverse moment $\lambda_B$, we will take a benchmark value $\lambda_B$ = 0.2 GeV for the phenomenological analysis in Section \ref{sec:analysis}, though it has an uncertainty not too small: its value is related to BR($B \to \gamma \ell \nu$), and when BR$(B \to \gamma \ell \nu) = 2.0 \times 10^{-6}$ with 20\% error is observed in future experiment, we will obtain $167 \, {\rm MeV} < \lambda_B < 304 \, {\rm MeV}$~\cite{Beneke:2011nf}. 
As for power corrections and perturbative corrections, the studies of $B \to \gamma \ell \nu$, see \textit{e.g.} \cite{Wang:2016beq,Wang:2018wfj,Beneke:2018wjp}, indicate that the modification to~\eqref{eq:GamBqtogammaZp} is not big and
thus will not change the main conclusion of our paper. Moreover, it is worth to note that the $m_{Z'}$ is not chosen to be too close to $M_{B_q}$ such that the photon energy $E_\gamma = |\boldsymbol{k}|$ is large enough for a converge power expansion of $1/E_\gamma$.

Interestingly, the result \eqref{eq:GamBqtogammaZp} depends not only on $g_V^{(q)}$ but also on $g_A^{(q)}$, which is different from the $B_q \to K Z'$ case. As a result, the branching ratio $B_q \to \gamma Z'$ suffers from both constraints from $B \to P K$ and $B \to V Z'$ where $P$ is a pseudoscalar meson and $V$ is a vector meson.

\subsection{The decay width of $Z'$}

If the $Z'$ boson also couples to the SM leptons through the following Lagrangian \footnote{In this Lagrangian, we assume that there are no flavor violating couplings in the lepton sector to avoid the severe constraints from lepton flavor violating (LFV) processes. },
\begin{align}
-{\cal L}_{Z'}^{\rm lep} = \bar{\ell} \gamma^\mu \left[ (g_L)_{\ell \ell} P_L +(g_R)_{\ell \ell} P_R \right] \ell Z'_\mu + (g_L)_{\nu \nu} \bar{\nu} \gamma^\mu P_L \nu Z'_\mu + h.c.,
\label{eq:LagZ'lep}
\end{align}
it can decay into lepton pairs when $m_{Z'}$ is larger than its mass threshold. 
The decay widths can be estimated as
\begin{align}
\Gamma (Z' \to \ell^- \ell^+) &= \frac{m_{Z'}}{24 \pi} \sqrt{1 - 4 r'_{\ell}} \Bigl\{ \left( |(g_L)_{\ell \ell}|^2 + |(g_R)_{\ell \ell}|^2 \right) ( 1 - r'_{\ell} ) + 6 {\rm Re} \! \left[ (g_L)_{\ell \ell} (g_R)_{\ell \ell}^{\ast} \right] r'_{\ell} \Bigr\}, \\
\Gamma (Z' \to \nu \bar{\nu}) &= \frac{m_{Z'}}{24 \pi} |(g_L)_{\nu \nu}|^2,
\end{align}
where $r'_{\ell} \equiv m_{\ell}^2 / m_{Z'}^2$, and the neutrino masses are neglected in $\Gamma (Z' \to \nu \bar{\nu})$. Note that when $Z'$ couples to the charged leptons with the vector-like form only, {\it i.e.}, $(g_L)_{\ell \ell} = (g_R)_{\ell \ell} \equiv g_{\ell}$, its decay width can be simplified as
\begin{align}
\Gamma (Z' \to \ell^- \ell^+) = \frac{m_{Z'}}{12 \pi} |g_{\ell}|^2 \sqrt{1 - 4 r'_{\ell}} \, ( 1 + 2 r'_{\ell}).
\end{align}

\section{Signatures of $B_q \to \gamma Z'$}
\label{sec:analysis}

In this section, we will consider the signatures of the decay $B_q \to \gamma Z'$. 
We choose a simplified case in which $Z'$ only has one decay channel, namely, one of the decay widths of $Z'$ saturates the total decay width of $Z'$: BR$(Z' \to \ell \bar{\ell}) \approx 1$ ($\ell = e, \mu$ or $\nu$). 
Therefore, the following analysis is independent on the $Z'$ coupling to the leptons. 
Furthermore as mentioned earlier, it is assumed that there are no flavor changing couplings in the lepton sector. 

\subsection{The channel $Z' \to \ET$}
\subsubsection{SM backgrounds}

In the SM, the background is originated from $B_q \to \bar{\nu} \nu \gamma$, mediated by the $W$-boson loop in the box and $Z$ penguin 
diagrams. Note that here the photon is radiated from an initial-state quark.
Without the photon, the decay will be extremely suppressed due to helicity conservation and similar conclusion applies to other channels 
$B \to e^+e^- \gamma$ and $B \to \mu^+\mu^- \gamma$. 
The branching ratio was first calculated in Ref.~\cite{Aliev:1996sk,Lu:1996et}, 
and the SM predictions are found in Ref.~\cite{Badin:2010uh} to be BR$(B_d \to \bar{\nu} \nu \gamma) = 1.96 \times 10^{-9}$ and BR$(B_s \to \bar{\nu} \nu \gamma) = 3.68 \times 10^{-8}$.

\subsubsection{Analysis: current $\&$ future}
\label{sec:Zpnnanalysis}

First of all, we collect the constraints from $B_q \to K Z'$. If the $Z'$ decays invisibly into neutrinos or dark sector particles, we should look for $B \to K + \ET$. 
The SM background $B \to K \nu \bar{\nu}$ must also be taken into account. 
The Belle II will search for three different decay modes $B^+ \to K ^+ \nu \bar{\nu}$, $B^0 \to K^{0*} \nu \bar{\nu}$ and $B^+ \to K^{+*} \nu \bar{\nu}$. 
Branching ratios to these channels have been searched by BaBar \cite{Lees:2013kla} and Belle \cite{Grygier:2017tzo,Lutz:2013ftz} experiments as shown in Table~\ref{T1}, where we also show the other relevant limits to the following analysis. 

\begin{table}[htb]
 \begin{center}
 \begin{tabular}{|c|c|c|c||c|c|c|c|} 
 \hline
 Decay mode & BaBar & Belle & Belle II & Decay mode & BaBar & Belle & Belle II \\ [0.5ex] 
 \hline
 $B^0 \to \pi^0 \nu \bar{\nu}$ & - & $< 0.9 \times 10^{-5}$ & - & $B^+ \to \pi^+ \nu \bar{\nu}$ & $< 1.0 \times 10^{-4}$ & $< 1.4 \times 10^{-5}$ & - \\ 
 \hline
 $B^0 \to \rho^0 \nu \bar{\nu}$ & - & $< 4.0 \times 10^{-5}$ & - & $B^+ \to \rho^+ \nu \bar{\nu}$ & - & $< 3.0 \times 10^{-5}$ & - \\
 \hline
 $B^0 \to K^0 \nu \bar{\nu}$ & $< 4.9 \times 10^{-5}$ & $< 2.6 \times 10^{-5} $ & - & $B^+ \to K^+ \nu \bar{\nu}$ & $< 1.6 \times 10^{-5}$ & $< 1.9 \times 10^{-5}$ & 11\% \\
 \hline
 $B^0 \to K^{*0} \nu \bar{\nu}$ & $< 1.2 \times 10^{-4}$ & $< 1.8 \times 10^{-5} $ & 9.6\% & $B^+ \to K^{*+} \nu \bar{\nu}$ & $< 6.4 \times 10^{-5}$ & $< 4.0 \times 10^{-5}$ & 9.3\% \\
 \hline
 \end{tabular}
\end{center}
\caption{Observed experimental limits on $B \to P \nu \bar{\nu}$ ($P = \pi, K$) and $B \to V \nu \bar{\nu}$ ($P = \rho, K^{*}$) decay branching ratios from BaBar \cite{Lees:2013kla,Aubert:2004ws} and Belle \cite{Grygier:2017tzo,Lutz:2013ftz}. The column of ``Belle II" shows the sensitivities on the branching ratio with 50 ab$^{-1}$ \cite{Kou:2018nap}.}
\label{T1}
\end{table}
In the SM, the decay branching ratios for these channels are estimated as BR$(B \to K^* \nu \bar{\nu}) = (9.6 \pm 0.9) \times 10^{-6}$ and BR$(B^+ \to K^+ \nu \bar{\nu}) = (4.6 \pm 0.5) \times 10^{-6}$ \cite{Kou:2018nap}. 
The Belle II will be able to observe these decay modes, and in particular the sensitivity to the branching ratio of $B \to K^{(*)} \nu \bar{\nu}$ will be about 10\% with 50 ab{$^{-1}$} \cite{Kou:2018nap}. 

\begin{figure}[thb]
  \begin{center}
    \includegraphics[width=0.49\textwidth,bb= 0 0 450 283]{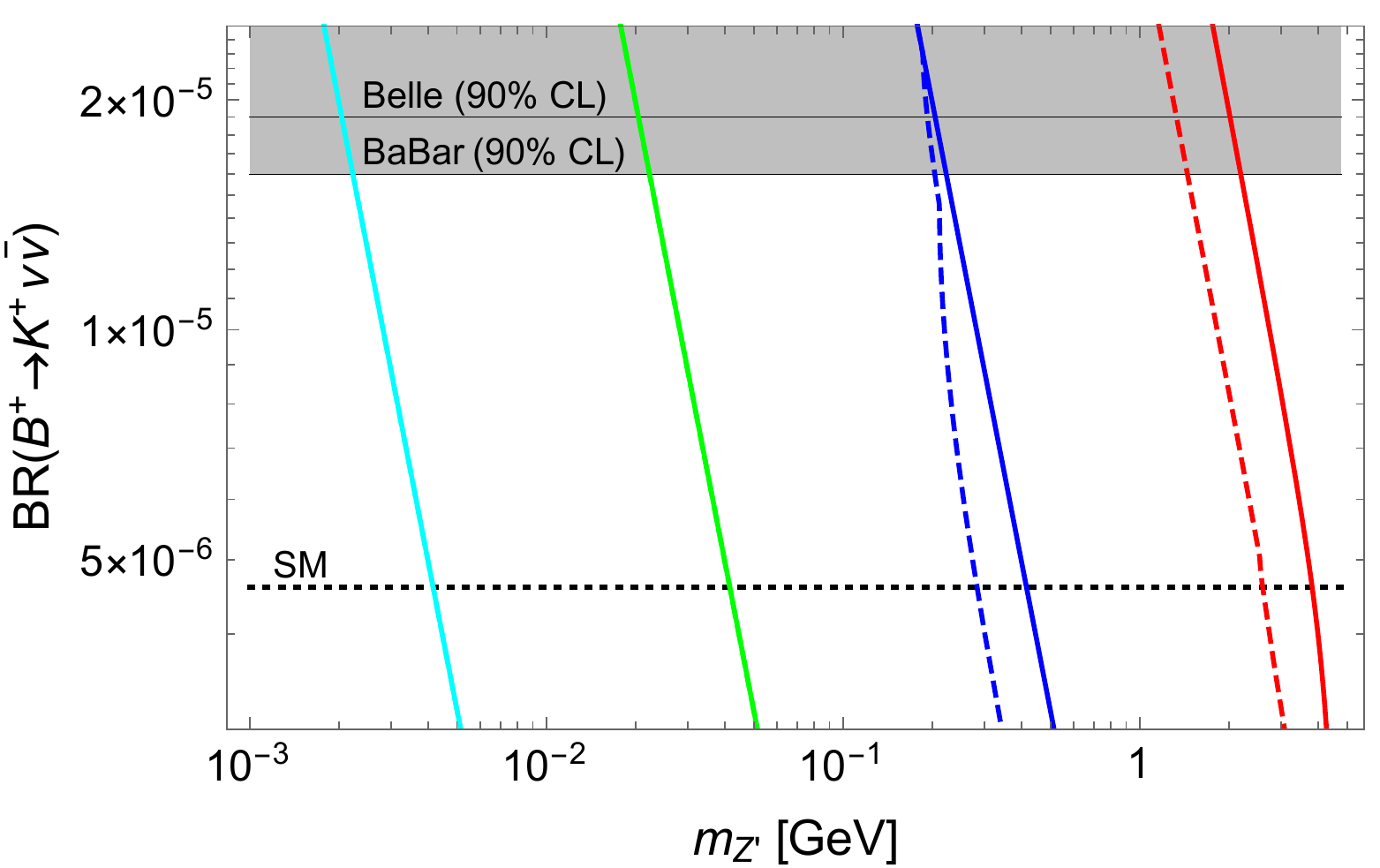}\hspace{0.5em}
    \includegraphics[width=0.49\textwidth,bb= 0 0 450 278]{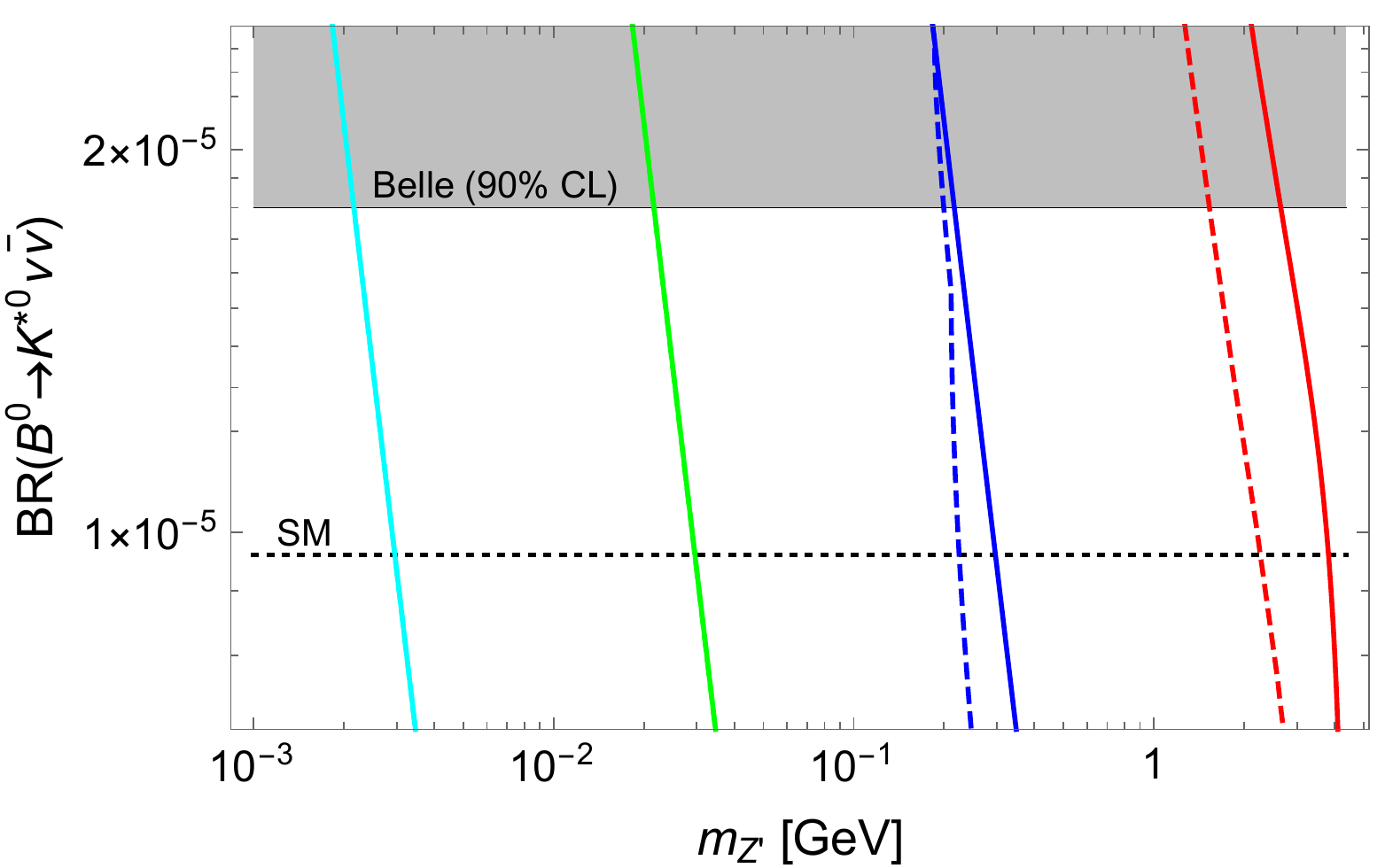}
    \caption{Branching ratios for $B^+ \to K^+ \nu \bar{\nu}$ (left) and $B^0 \to K^{*0} \nu \bar{\nu}$ (right). The horizontal black lines are current limits from the BaBar and Belle, and shaded regions are excluded by these experiments. The dashed black lines show the SM predictions. Red, blue, green and cyan lines are our NP predictions with $(g_L)_{sb} = 10^{-8}$, $10^{-9}$, $10^{-10}$ and $10^{-11}$. $(g_R)_{sb} = (g_L)_{sb}$ in the left panel, and $(g_R)_{sb} = - (g_L)_{sb}$ in the right panel. The dashed red and blue lines show the predictions for the model in Ref.~\cite{Kang:2019vng}.}
    \label{fig:BtoKnunu}
  \end{center}
\end{figure}
We display the current limits and our predictions for $B \to K \nu \bar{\nu}$ decay in Fig.~\ref{fig:BtoKnunu}. 
The dashed black lines show the central values of the SM predictions. 
The horizontal black lines denote the current limits from BaBar and Belle, and the shaded regions have been excluded by these experiments. 
Recall that the $B$ decays into the pseudoscalar/vector $K$ mesons are sensitive to the vector/axial-vector couplings $g_V^{(s)}/g_A^{(s)}$. 
For each type, in this figure we show the stronger constraints between the $B$ and $B^+$ decays collected in Table~\ref{T1}. 
Concretely, $B^+ \to K^+ \nu \bar{\nu}$ and $B^0 \to K^{*0} \nu \bar{\nu}$ are chosen in the left and right panels, respectively. 
Then, the red, blue, green and cyan lines correspond to our NP predictions with $(g_L)_{sb} = 10^{-8}$, $10^{-9}$, $10^{-10}$ and $10^{-11}$, respectively.
We also show the predictions for the model of Ref.~\cite{Kang:2019vng} in dashed red and blue lines.

From this figure, the branching ratio can exceed the SM background in a wide range of $m_{Z'}$ by choosing the vector or axial-vector coupling. 
We obtain the following constraints on $g_{V,A}^{(s)}$ in each $m_{Z'}$ value:
\begin{align}
| g_V^{(s)} | &\lesssim 9.2 \times 10^{-9} \left( \frac{m_{Z'}}{\rm GeV} \right) ~~~ \text{from} ~ {\rm BR}(B^+ \to K^+ \nu \bar{\nu}), \label{eq:congVs} \\[0.3ex]
| g_A^{(s)} | &\lesssim 8.6 \times 10^{-9} \left( \frac{m_{Z'}}{\rm GeV} \right) ~~~ \text{from} ~ {\rm BR}(B^0 \to K^{*0} \nu \bar{\nu}). \label{eq:congAs}
\end{align}
We have assumed approximate scaling relations for the decay widths, e.g., $\propto g^{(s)}_V/m_{Z'}$ for $\Gamma(B \to K Z')$, which is seen in Eq.~(\ref{BtoK}) after neglecting the phase space factor, reasonable for $m_{Z'} < 2 \mathchar`- 3$ GeV. 
However for $m_{Z'}$ around 3-4 GeV, the phase space suppression becomes significant, and then these relations are invalid. 
In general, by combining these constraints, the flavor violating couplings $(g_{L,R})_{sb}$ should be smaller than $10^{-8}$ when $m_{Z'} < \mathcal{O}(1)$ GeV.

We also calculate the bounds on $(g_{L,R})_{db}$ from similar decay channels. 
The current experimental constraints for $B \to \pi \nu \bar{\nu}$ and $B \to \rho \nu \bar{\nu}$ are shown in Table~\ref{T1}, and they give the following bounds
\begin{align}
| g_V^{(d)} | &\lesssim 1.1 \times 10^{-8} \left( \frac{m_{Z'}}{\rm GeV} \right) ~~~ \text{from} ~ {\rm BR}(B^+ \to \pi^+ \nu \bar{\nu}), \label{eq:congVd} \\[0.3ex]
| g_A^{(d)} | &\lesssim 1.3 \times 10^{-8} \left( \frac{m_{Z'}}{\rm GeV} \right) ~~~ \text{from} ~ {\rm BR}(B^+ \to \rho^+ \nu \bar{\nu}), \label{eq:congAd}
\end{align}
when $m_{Z'} < 4$ GeV for BR$(B^+ \to \pi^+ \nu \bar{\nu})$ and $m_{Z'} < 2$ GeV for BR$(B^+ \to \rho^+ \nu \bar{\nu})$. 
Note that constraints from the $B_q \mathchar`- \overline{B}_q$ mixing are much weaker than the above constraints in Eqs.~\eqref{eq:congVs}-\eqref{eq:congAd}. 
Therefore, we ignore them in the following analyses.

Now we move to the new decay channel $B_q \to \gamma+ Z'( \to \ET)$. 
Currently, no measurement for $B_s \to \gamma + \ET$ decay is available yet, while $B_d \to \gamma + \ET$ has been searched by BaBar~\cite{Lees:2012wv,Tanabashi:2018oca} and an upper limit on its branching ratio is given:
\be
{\rm BR}(B_d \to \gamma + \ET) < 1.7 \times 10^{-5}\,\, .
\label{limit1}
\ee
Recently, the Belle reported the updated result for this decay mode \cite{Ku:2020aix}, and an upper limit as BR$(B_d \to \gamma + \ET) < 1.6 \times 10^{-5}$ is obtained.
Note that in the present framework we interpret $\ET$ as the invisible neutrino pair from $Z' \to \bar\nu\nu$, but as stated before, it may also be a pair of low mass dark matter candidate $X$ and even the $Z'$ itself when it behaves as a stable particle at the detector.

\begin{figure}[tb]
  \begin{center}
    \includegraphics[width=0.45\textwidth,bb= 0 0 420 427]{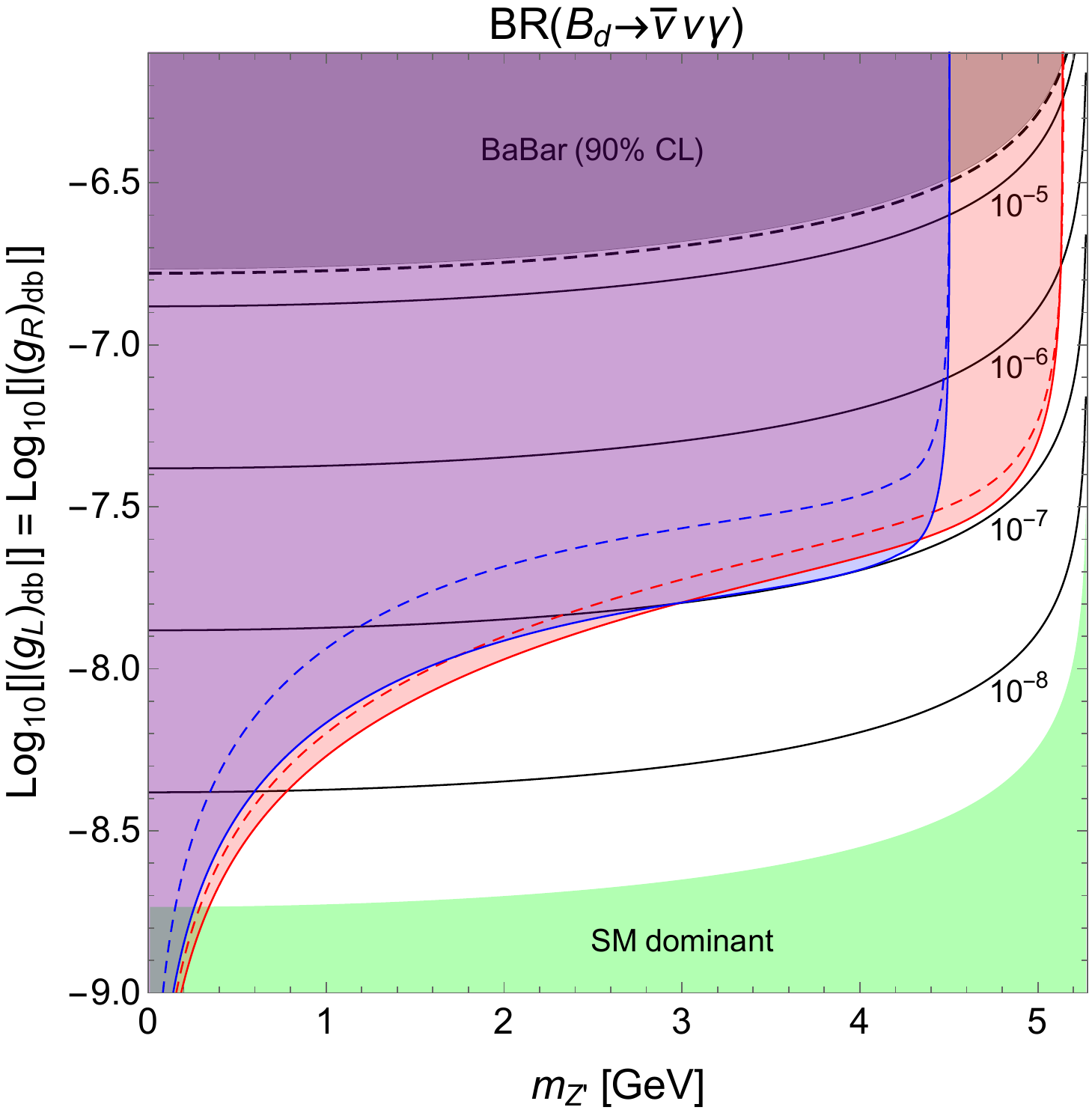} \hspace{1.0em}
    \includegraphics[width=0.45\textwidth,bb= 0 0 420 427]{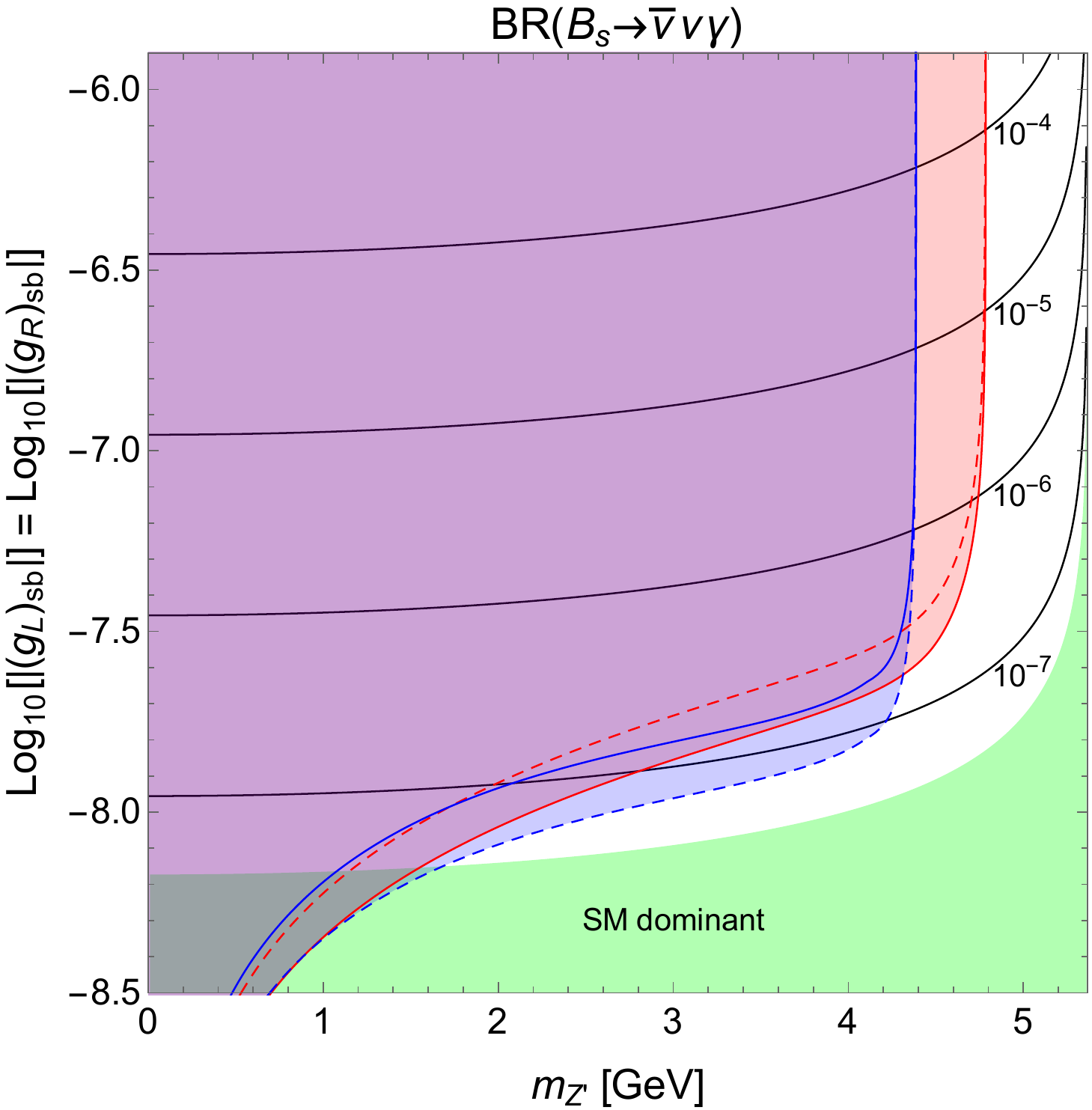}
    \caption{Branching ratios of $B_q \to \gamma + \ET$. The gray shaded region is the direct bound from BaBar \cite{Lees:2012wv}, and the green shaded region shows the area where the prediction is smaller than the SM prediction. Black dashed line shows the direct bound from Belle \cite{Ku:2020aix}. Red and blue shaded regions show the experimental bounds from $B \to P \nu \bar{\nu}$ ($P = \pi$ for left panel and $P = K$ for right panel) and $B \to V \nu \bar{\nu}$ ($V = \rho$ for left panel and $V = K^*$ for right panel), respectively. These bounds involving $P^+$ or $V^+$ ($P^0$ or $V^0$) are shown by red or blue solid (dashed) lines. Note that with $m_{Z'} > $ 4 GeV, our calculation is not concrete.}
    \label{fig:BtogamMET}
  \end{center}
\end{figure}
\begin{figure}[hbt]
  \begin{center}
    \includegraphics[width=0.449\textwidth,bb= 0 0 400 401]{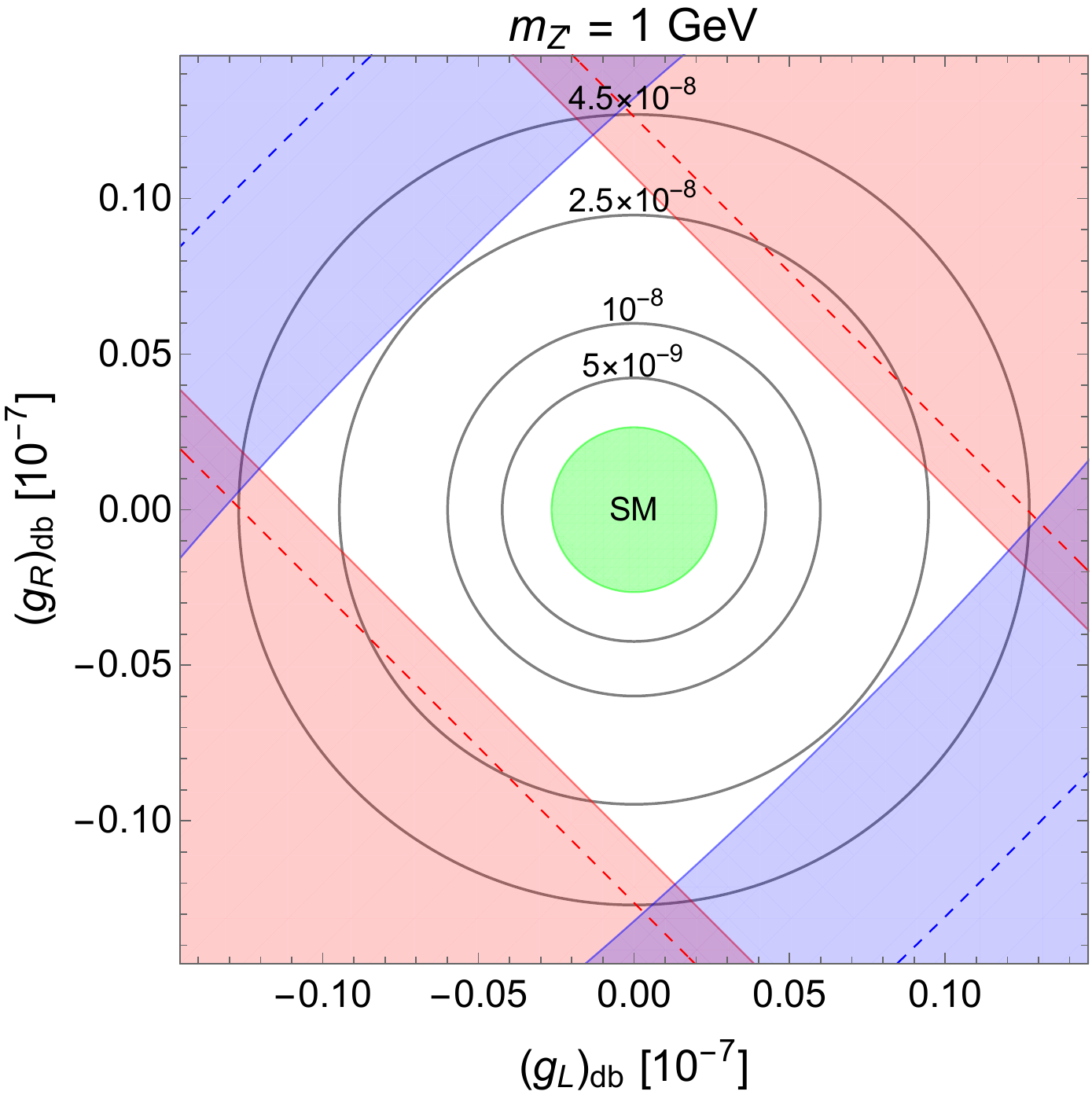} \hspace{1.0em}
    \includegraphics[width=0.44\textwidth,bb= 0 0 400 409]{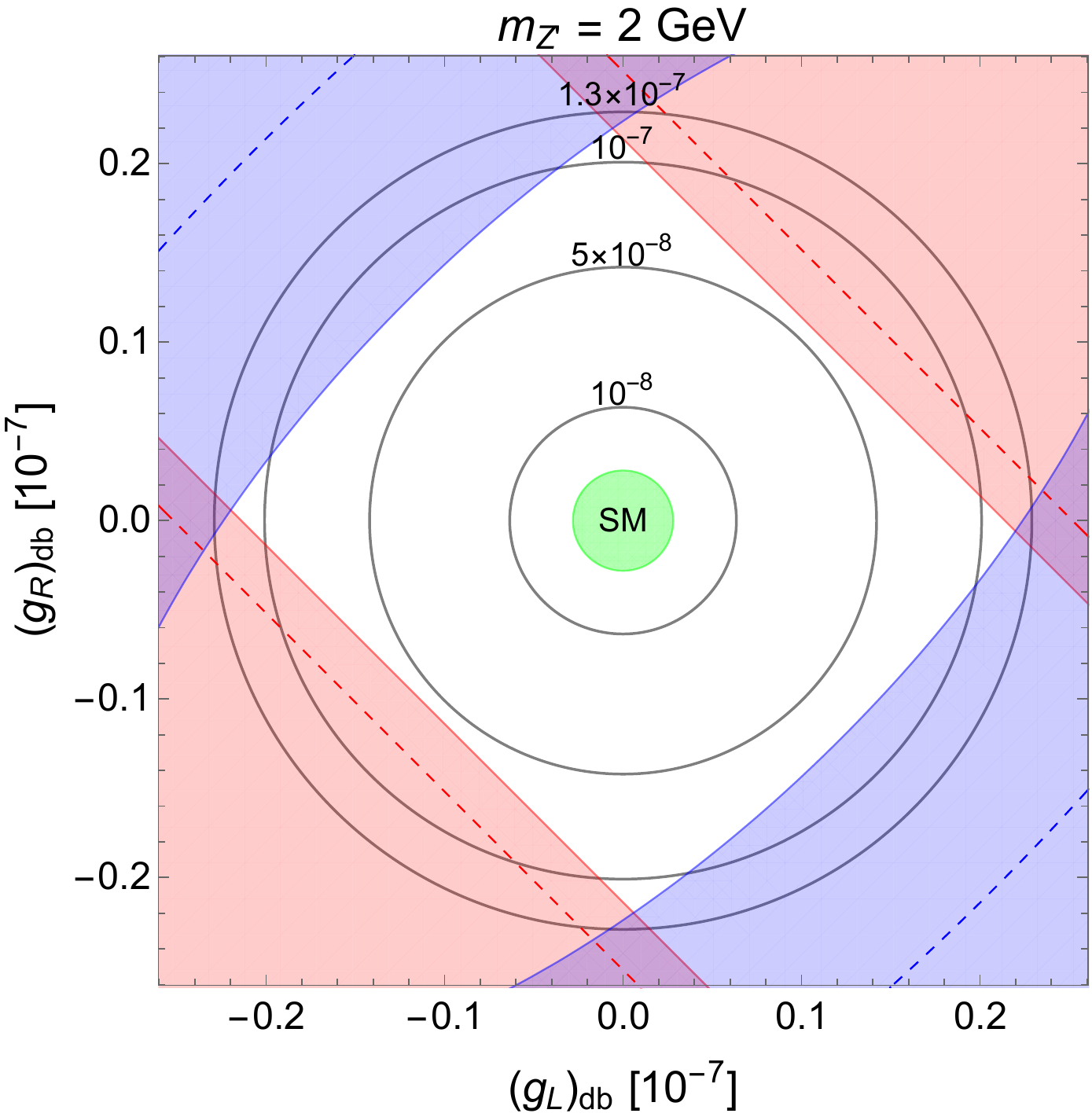} \\[1.5ex]
    \includegraphics[width=0.44\textwidth,bb= 0 0 400 409]{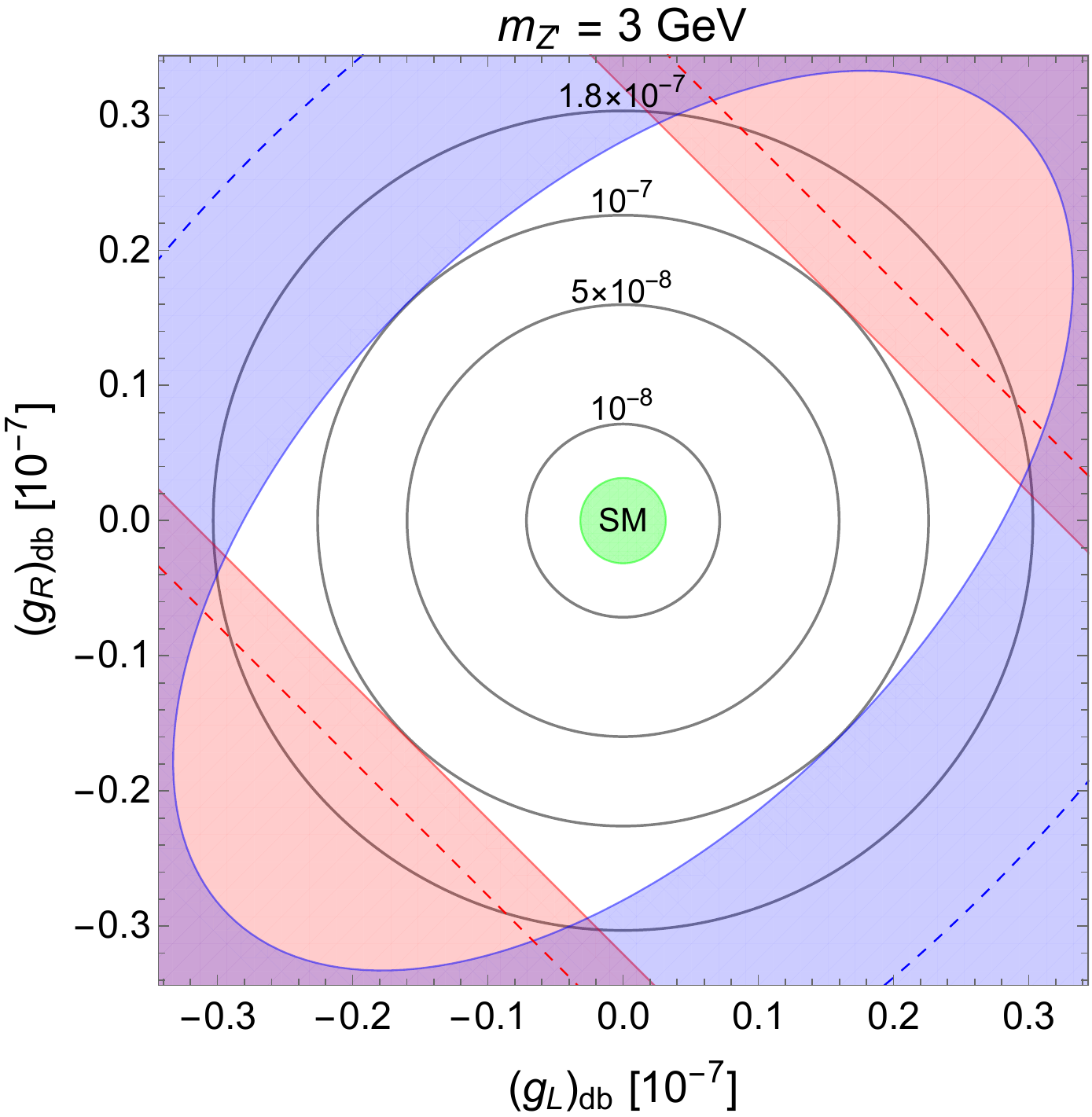} \hspace{1.0em}
    \includegraphics[width=0.44\textwidth,bb= 0 0 400 409]{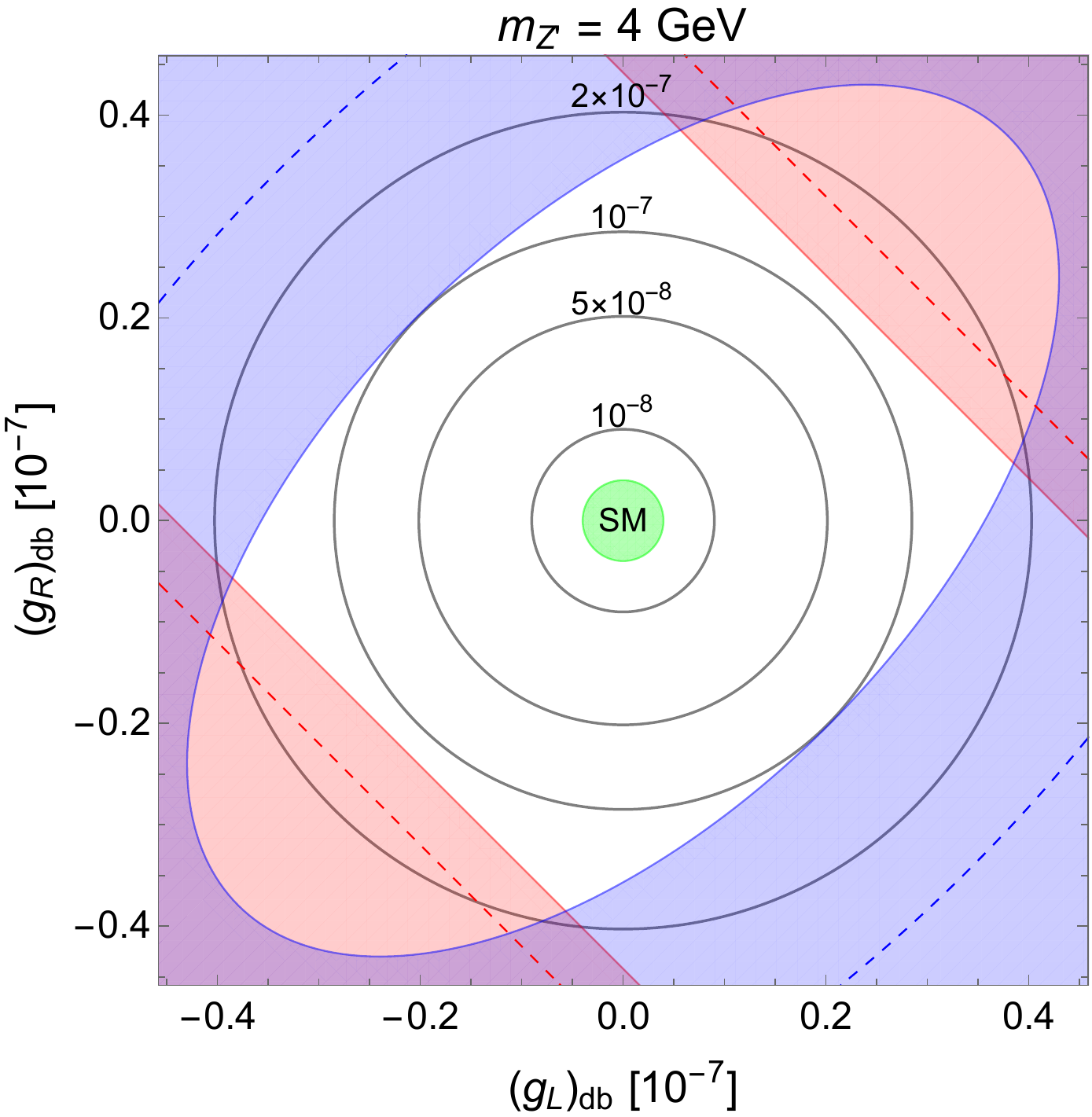}
    \caption{Branching ratios of $B_d \to \gamma + Z'(\to \ET)$ with fixed $m_{Z'}$. The meanings of each color plot are the same as Fig.~\ref{fig:BtogamMET}. The region denoted as ``SM" shows the region where the prediction is smaller than the SM prediction.}
    \label{fig:BRBdfixmZp}
  \end{center}
\end{figure}

Before the demonstration of the results of BR($B_q \to \gamma+Z'(\to \ET)$), we would like to explain the way to present them. Besides the $m_{Z'}$ dependence, the branching ratios depend on both $g_V^{(s)}$ and $g_A^{(s)}$, which are supposed to be independent free parameters in our effective theory. Considering that their allowed values are confronting with the strong exclusions from BR$(B \to P \nu \bar{\nu})\propto( g_V^{(q)})^2$ and BR$(B \to V \nu \bar{\nu})\propto( g_A^{(q)})^2$, it is illustrative to consider two limiting cases with $(g_L)_{qb} = \pm (g_R)_{qb} $, which reduces one parameter. Then we show the results in the $(m_{Z'}, \log [ | (g_L)_{qb} | ])$ plane in Fig.~\ref{fig:BtogamMET}. Later, in Fig.~\ref{fig:BRBdfixmZp} we will choose several benchmark values for $m_{Z'}$ and display the results in the $( (g_L)_{db}, (g_R)_{db} )$ plane.

In Fig.~\ref{fig:BtogamMET}, the gray shaded region in the left panel denotes for the direct experimental bound on the signature $B_q \to \gamma+Z'(\to \ET)$ from BaBar \cite{Lees:2012wv}, using the data Eq.~(\ref{limit1}), and the black dashed line shows the direct bound from recent Belle result \cite{Ku:2020aix}; these bounds are quite weak compared to the indirect bounds. 
The red and blue shaded regions show the indirect bounds from the decays of $B$ into the pseudoscalar meson $P$ and into the vector meson $V$, respectively. 
Here, $P = \pi$ and $V = \rho$ in the left panel, while $P = K$ and $V = K^*$ in the right panel \footnote{Although there is another experimental limit from $B_s \to \phi \nu \bar{\nu}$ \cite{Adam:1996ts}, its current limit on $(g_{L,R})_{sb}$ is weaker than that of $B \to K \nu \bar{\nu}$. Therefore, we omit it in the right panel of Fig.~\ref{fig:BtogamMET}. }. 
In the vector coupling limit (corresponding to the red region) or axial-vector coupling limit (corresponding to the blue region) under consideration, only the appropriate bound does exist, but in general they coexist. 
Moreover, since the bounds come from both the charged states $P^+/V^+$ and the neutral states $P^0/V^0$, to distinguish them we label the boundaries as solid lines for the former while dashed lines for the latter. 
In the green shaded region, the NP contribution to $B_q \to \gamma+Z'(\to \ET)$ is smaller than the SM background.

From the right panel, one can see that $B \to K^{*} \nu \bar{\nu}$ already give fairly severe constraints, and consequently they squeeze the room for a sizable NP contribution to BR($B_s \to \gamma+Z'(\to \ET)$). 
Of course, there is still a wide room in the relatively heavy-$Z'$ region, that is $M_{B_s} - M_{K^{*}} < m_{Z'} < M_{B_s}$ with $M_{K^{*}}\approx 892$ MeV, where the two body decay $B_s \to K^*+Z'$ is kinematically forbidden. 
However, with $m_{Z'} > $ 4 GeV, our calculation is not concrete as mentioned in Section~\ref{sec:calcBgZp}. 
On the other hand, from the left panel, the branching ratio can exceed the SM background by two orders of magnitude, $\sim \mathcal{O}(10^{-7})$, even when $m_{Z'} = \mathcal{O}(1)$ GeV. 
Therefore, if the future Belle II can search the branching ratio of BR($B_d \to \gamma+Z'(\to \ET)$) with size of $\mathcal{O}(10^{-7})$ level, there is a possibility to find new physics effects in this decay mode, although this is about two orders of magnitude smaller than the current experimental bound from the BaBar and Belle.

We also show the prospects of the signatures, to be specific $B_d \to \gamma + Z'(\to \ET)$ for a better opportunity to accommodate NP, in the $( (g_L)_{db}, (g_R)_{db} )$ plane in Fig.~\ref{fig:BRBdfixmZp} by choosing several values of $m_{Z'}$. Concretely, we take $m_{Z'} = 1$ GeV (top left), $2$ GeV (top right), $3$ GeV (bottom left) and $4$ GeV (bottom right) as examples. 
In this demonstration, we show the impact of the structure of the $Z'$ couplings on the NP signature. 
In general, it does not make a big difference to the limiting analysis made before. 
This is well expected because $B_d^+ \to \pi^+ \nu \bar{\nu}$ and $B_d^+ \to \rho^+ \nu \bar{\nu}$ give compensatory constraints on $g_V^{(d)}$ and $g_A^{(d)}$. 
However, a mild enhancement of the allowed BR($B_d \to \gamma + Z'(\to \ET)$) is still possible. 
For example, from Fig.~\ref{fig:BtogamMET} it is seen that for $m_{Z'} = 3$ GeV, its allowed value should lie below $10^{-7}$, whereas here it reaches $1.7 \times 10^{-7}$ for $(g_L)_{db} = 0.29 \times 10^{-7}$ and $(g_R)_{db} = 0.03 \times 10^{-7}$.

\subsection{The channel $Z'\to e^+e^-$}
Apart from the invisible decay modes, $Z'$ can also decay visibly such as an electron pair and generates the signature $B_q \to e^+e^- \gamma$. 
Of course, the possible size of NP is restricted by $B \to V/P+e^+e^- $. 
As a matter of fact, this rare decay signature, along with the one $B \to V/P+\mu^+\mu^- $ discussed in the following subsection, is of special interests in the recent years, since it hints violation of the lepton universality in the SM.

\subsubsection{$R_{K^*}$ anomaly and an FCNC light $Z'$}
\label{RK}

Before the discussions of our signature, let us briefly review this anomaly and how a light $Z'$ described by the effective model Eq.~(\ref{eq:LagZ'}) could play an essential role in resolving the anomaly. Lepton flavor universality is measured by ratios such as $R_{K}=\frac{{\rm BR}(B \to K \mu^+\mu^-)}{{\rm BR}(B \to K e^+e^-)}$. Recently, the LHCb collaboration determined~\cite{Aaij:2014ora,Aaij:2017vbb,Aaij:2019wad}
\begin{eqnarray*}
R_{K^{*0}} = 
\begin{cases}
0.66^{+0.11}_{-0.07}\mathrm{\,(stat)} \pm 0.03\mathrm{\,(syst)}	& \textrm{for } 0.045 < q^{2} < 1.1~\mathrm{\,Ge\kern -0.1em V^2\!/}c^4 \, , \\
0.69^{+0.11}_{-0.07}\mathrm{\,(stat)} \pm 0.05\mathrm{\,(syst)}	& \textrm{for } 1.1\phantom{00} < q^{2} < 6.0~\mathrm{\,Ge\kern -0.1em V^2\!/}c^4 \, ,
\end{cases}
\label{LHCb}
\end{eqnarray*}
and 
\begin{eqnarray*}
R_{K} &= 0.745^{+0.090}_{-0.074}\mathrm{\,(stat)} \pm 0.036\mathrm{\,(syst)}~~	& \textrm{for } 1 < q^{2} < 6~\mathrm{\,Ge\kern -0.1em V^2\!/}c^4 \, \textrm{(Run-1)} \, , \label{LHC2} \\
R_{K} &= 0.846^{+0.060}_{-0.054}\mathrm{\,(stat)} {}^{+0.016}_{-0.014}\mathrm{\,(syst)}~~~~~~	& \textrm{for } 1.1 < q^{2} < 6~\mathrm{\,Ge\kern -0.1em V^2\!/}c^4 \, \textrm{(Run-2)} \, , \label{LHC13TeV}
\end{eqnarray*}
with $q^2$ the dilepton invariant mass squared. 
The combined results hint lepton flavor non-university. By adding new heavy FCNC particles fails in explaining them simultaneously, in particular the low $q^2$ bin data $R_{K^*}^{\rm low}$. Whereas a light $Z'$ ($m_{Z'} \lesssim 2 m_{\mu}$)~\footnote{When $2 m_{\mu} < m_{Z'} < m_B$, $Z'$ contributions should be observed as a resonance in the dimuon invariant mass if the $Z'$ width is narrow. Such signatures are not observed, and hence, the mass range which has a possibility to explain the anomaly is $m_{Z'} \lesssim 2 m_{\mu}$ or $m_B \lesssim m_{Z'}$.} as in Eq.~(\ref{eq:LagZ'}), and moreover coupling to electron provides a solution~\cite{Ghosh:2017ber,Datta:2017ezo,Altmannshofer:2017bsz}; it generates non-local operator like $(\bar s \gamma^\mu P_L b)(\bar e \gamma_\mu e)$ whose Wilson coefficient, unlike in the heavy new physics, is $q^2$-dependent.

Actually, before the LHC running, the Belle has already performed such a search \cite{Wei:2009zv}. They reported the individual branching ratios of ${\rm BR}(B \to P/V \ell^+ \ell^-)$, which are listed in Table~\ref{T2} and Table~\ref{T3}. The resulting $R_K$ and $R_{K^*}$ are given by
\begin{eqnarray*}
R_{K} = 1.03~{\pm~0.19}\mathrm{\,(stat)} \pm 0.06\mathrm{\,(syst)} \, , \\
R_{K^*} = 0.83~{\pm~0.17}\mathrm{\,(stat)} \pm 0.08\mathrm{\,(syst)}\, . 
\label{belle}
\end{eqnarray*}
It is expected that Belle II can measure $R_K$ with less than 5\% uncertainty for 50 ab{$^{-1}$}~\cite{Kou:2018nap}, to double check this anomaly. The light resonance solution based on Eq.~(\ref{eq:LagZ'}) gives rise to the signature $B_s \to \gamma + Z'(\to e^+e^-)$, and thus it is of interest to investigate if there is a possibility to cross check that solution.

\subsubsection{SM backgrounds}

In the SM, $B_q \to e^+e^- \gamma$ is produced similarly to the previous case $B_q \to \bar{\nu} \nu \gamma$, from the $W$-box and $Z$-penguin diagrams. Likewise, the $\gamma$-emission is helpful to overcome the helicity suppression. The decay branching ratio for $B_q\to e^+e^- \gamma$ in the SM was first calculated by Aliev {\it et al.} in the framework of light-cone QCD sum rule \cite{Aliev:1996ud}, obtaining BR$(B_d \to e^+e^- \gamma) = 1.5 \times 10^{-10}$ and BR$(B_s\to e^+e^- \gamma)=2.35\times 10^{-9}$. In a recent work Kozachuk {\it et al.} performed a revised calculation and they obtained a smaller branching ratio for $B_d$ while larger value for $B_s$:
\begin{align*}
{\rm BR}(B_d\to \ell^+ \ell^- \gamma)&=(1.05\pm 0.15)\times 10^{-11}\,,\\
\nonumber
{\rm BR}(B_s\to \ell^+ \ell^- \gamma)&=(6.01\pm 0.08)\times 10^{-9}\,,
\label{emu}
\end{align*}
with $\ell=e, \mu$ \cite{Kozachuk:2017mdk,Kozachuk:2018dqc}.

\subsubsection{Analysis}

We look for new physics signatures from $B_q \to \gamma + Z'(\to e^+e^- )$ decay process. The BaBar collaboration has searched for the decay signature $B_q \to e^+e^- \gamma$~\cite{Aubert:2007up,Tanabashi:2018oca}, but for $q=d$ only, and they obtained the upper limit for this decay branching ratio 
\be
{\rm BR}(B_d\to e^+e^- \gamma ) < 1.2 \times 10^{-7}\,\, .
\label{limit2}
\ee 
It is far above the SM prediction which is around the level of $10^{-11}$. Besides this direct bound, in Table~\ref{T2} we summarize the indirect bounds from BR($B \to P/V e^+ e^-$) \footnote{See also EPAPS Document No.~E-PRLTAO-103-030943 and EPAPS Document No.~E-PRLTAO-102-060910.}. Compared to the invisible mode, the visible mode gives a stronger (direct and indirect) bound by about two orders of magnitude. Note that 
the branching ratios for the modes $B \to P/V \ell^+ \ell^-$ with $\ell=e$ and also $\mu$ given later take measured values rather than upper bounds. As a simplifying discussion, we set a ``rough bound" by assuming that the new physics contribution does not exceed 30\% of each central value of the experimental results. Taking a serious statistical analysis will not change our discussions much. 

\begin{table}[htb]
 \begin{center}
 \begin{tabular}{|c|c|c||c|c|c|} 
 \hline
 Decay mode & BaBar & Belle & Decay mode & BaBar & Belle \\ [0.5ex] 
 \hline
 $B^0 \to \pi^0 e^+e^-$ & $< 8.4 \times 10^{-8}$ & $< 2.3 \times 10^{-7}$ & $B^+ \to \pi^+ e^+e^-$ & $< 1.25 \times 10^{-7}$ & $< 8.0 \times 10^{-8}$ \\ 
 \hline
 $B^0 \to K^0 e^+e^-$ & $0.8^{+1.5}_{-1.2} \pm 0.1$ & $2.0^{+1.4}_{-1.0} \pm 0.1$ & $B^+ \to K^+ e^+e^-$ & $5.1^{+1.2}_{-1.1} \pm 0.2$ & $5.7^{+0.9}_{-0.8} \pm 0.3$ \\
 \hline
 $B^0 \to K^{*0} e^+e^-$ & $8.6^{+2.6}_{-2.4} \pm 0.5$ & $11.8^{+2.7}_{-2.2} \pm 0.9$ & $B^+ \to K^{*+} e^+e^-$ & $13.8^{+4.7}_{-4.2} \pm 0.8$ & $17.3^{+5.0}_{-4.2} \pm 2.0$ \\
 \hline
 \end{tabular}
\end{center}
\caption{Observed experimental results on $B \to P/V e^+e^-$ decay branching ratios from BaBar \cite{Aubert:2008ps,Lees:2013lvs} and Belle \cite{Wei:2009zv,Wei:2008nv}. The results for $B \to K^{(*)} e^+e^-$ are shown in units of $10^{-7}$.}
\label{T2}
\end{table}

\begin{figure}[tb]
  \begin{center}
    \includegraphics[width=0.45\textwidth,bb= 0 0 420 427]{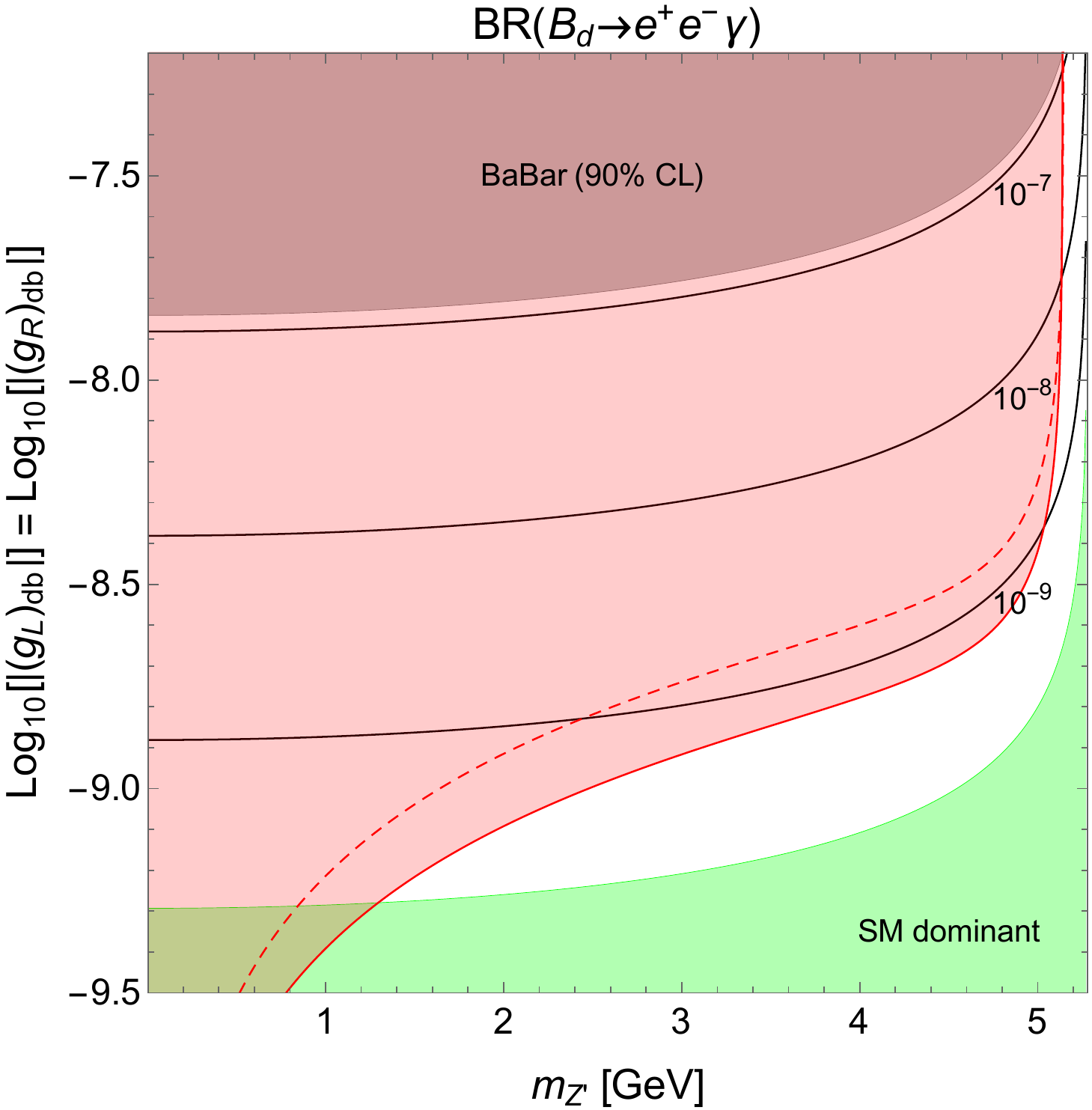} \hspace{1.0em}
    \includegraphics[width=0.45\textwidth,bb= 0 0 420 427]{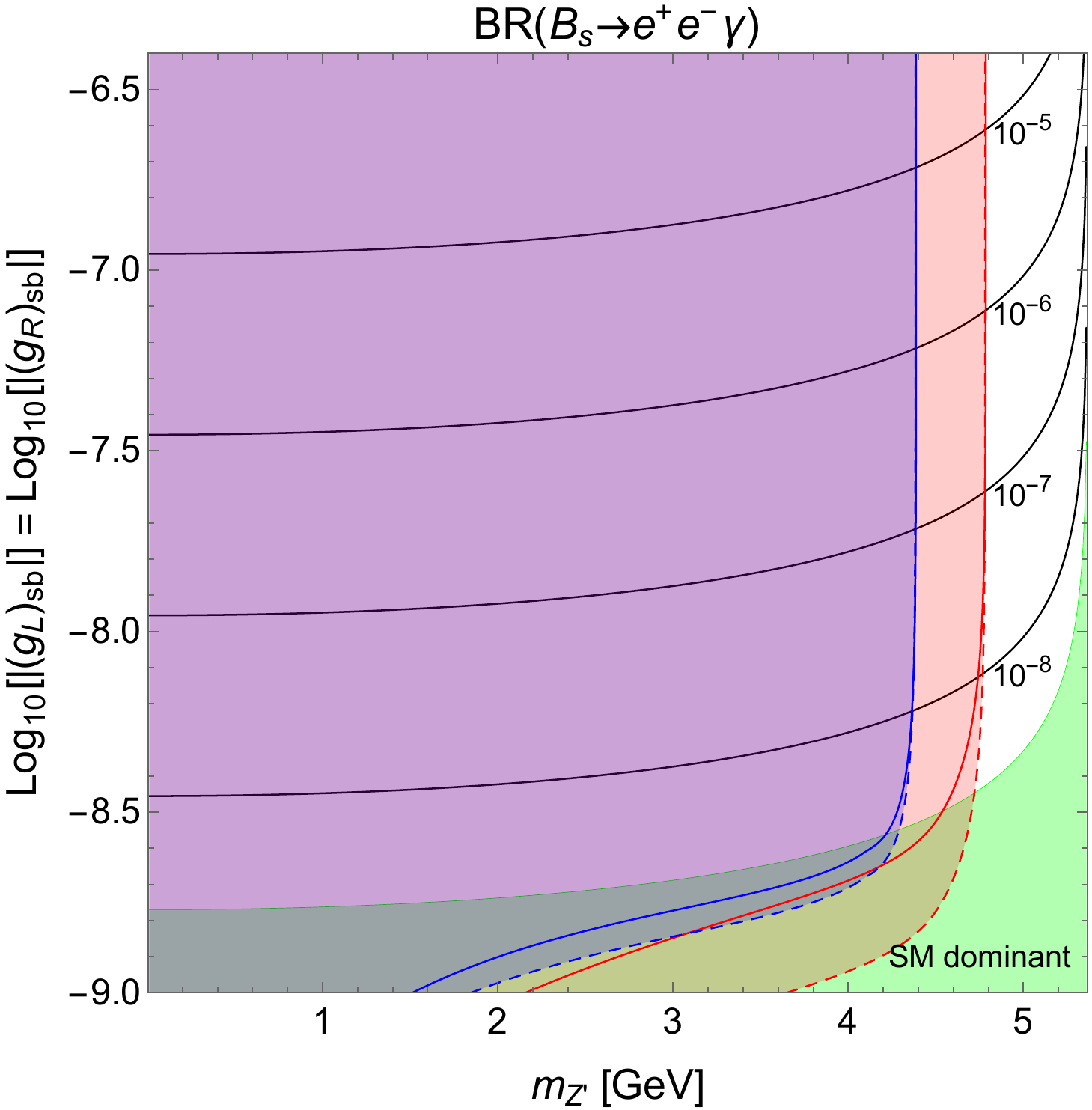}
    \caption{Branching ratios of $B_q \to e^+e^- \gamma$. Left panel is for $B_d \to e^+e^- \gamma$, and right panel is for $B_s \to e^+e^- \gamma$. The meanings of each color plot are the same as Fig.~\ref{fig:BtogamMET}.}
    \label{fig:Btogamee}
  \end{center}
\end{figure}

We show the results in Fig.~\ref{fig:Btogamee}, where 
the left panel is for $B_d \to \gamma + Z'(\to e^+e^- )$ and the right panel is for $B_s \to \gamma + Z'(\to e^+e^- )$. The meanings of each color plot are the same as Fig.~\ref{fig:BtogamMET}. In this case, the bounds come from $B \to P/V e^+e^-$, 
which are shown in Table~\ref{T2}. One point deserves special attention: Currently, there are no experimental constraints on $B \to V e^+e^-$ caused by $b \to d$ transition yet~\footnote{In the $\bar\nu\nu$ mode, the Belle measurement of $B^0 \to \rho^0 \nu \bar{\nu}$ constrains the axial-vector FCNC coupling $\bar b \gamma^5\gamma^\mu d Z'_\mu$. Nevertheless, currently there are no data of $B^0 \to \rho^0 \ell^+ \ell^-$, because they are theoretically unclean owing to the $\gamma$ penguin in addition to the $Z$ penguin contribution.}. Therefore, we only show the bound from $B_d \to \pi e^+e^-$ in the left panel. This means that if we are working in the axial-vector coupling limit $(g_L)_{db} = - (g_R)_{db}$, then BR$(B_d \to \gamma + Z'(\to e^+e^- ))$ is only constrained by the direct bound; it can be as large as $\sim {\cal O}(10^{-7})$, and thus we can expect that there is some new physics effect in $B_d \to e^+e^- \gamma$. 

By contrast, for $B_s \to \gamma + Z'(\to e^+e^- )$ caused by $b \to s$ one encounters the tight compensatory ``rough bound" from $B_d \to K^{*} e^+e^-$. Consequently, in the vector or axial-vector coupling limits, the allowed size of new physics contributions to the BR($B_s \to \gamma + Z'(\to e^+e^- )$) lies below the SM background except for near the $B$ meson threshold; see the right panel of Fig.~\ref{fig:BtogamMET}. Even if we relax the ``rough bound" to allow the new physics contribution equal to the central value of each experimental value, the maximal value of BR($B_s \to \gamma + Z'(\to e^+e^- )$) is comparable to the SM prediction: e.g., a value $\simeq 3 \times 10^{-9}$ for $m_{Z'} = 3$ GeV. Although there is a possibility to enhance its branching ratio by choosing $(g_{L,R})_{sb}$ as we shown in Fig.~\ref{fig:BRBdfixmZp}, we cannot expect a sizable enhancement. Actually, we find that the result is smaller than the SM prediction in any set of $((g_L)_{sb}, (g_R)_{sb})$ in the allowed region of above ``rough bound".

We end up this subsection with the answer to the question raised at the end of Section~\ref{RK}. According to Ref.~\cite{Altmannshofer:2017bsz}, the best fit to explain $R_{K^*}^{\rm low}$ requires BR$(B \to K^{\ast} Z') = \mathcal{O}(10^{-7})$ with $m_{Z'} \simeq 200$ MeV. For the model Eq.~(\ref{eq:LagZ'}), the value of this branching ratio corresponds to the one obtained from the couplings $(g_{L,R})_{sb}$ just below the indirect bounds in the right panel of Fig.~\ref{fig:Btogamee}. In other words, when we choose the appropriate size of couplings $(g_{L,R})_{sb}$ to explain $R_{K^*}^{\rm low}$, the effect of $Z^{\prime}$ to $B_s \to e^+ e^- \gamma$ is negligible compared with the SM contribution. Therefore, we cannot expect any signature of $B_s \to e^+ e^- \gamma$ for the model with the explanation of the $R_{K^*}^{\rm low}$ anomaly.

\subsection{The channel $Z' \to \mu^+\mu^-$}

\subsubsection{SM backgrounds}
Similar to the case of $B_q \to e^+e^- \gamma$, if $m_{Z'} \geq 2 m_{\mu}$, there 
will be new physics contributions to the $B_q \to \mu^+\mu^- \gamma$ channel 
where $Z'$ decays into $\mu^+ \mu^-$ pair. The SM contribution for this process 
is also calculated by Aliev {\it et al.} \cite{Aliev:1996ud}, and the corresponding 
decay branching ratios are BR$(B_d \to \mu^+\mu^- \gamma) = 1.2 \times 10^{-10}$ and BR$(B_s \to \mu^+\mu^- \gamma) = 1.9 \times 10^{-9}$. The recent calculations for the above 
channels in the SM by Kozachuk {\it et al.}~\cite{Kozachuk:2017mdk,Kozachuk:2018dqc} are already mentioned. Dubnicka {\it et al.} also provided an estimate for BR$(B_s \to \mu^+\mu^- \gamma) = 1.6 \times 10^{-9}$
in the SM using covariant quark model \cite{Dubnicka:2018gqg}.

\subsubsection{Analysis}
Experimental measurement of $B_d \to \mu^+\mu^-\gamma$ was also done by BaBar 
\cite{Aubert:2007up,Tanabashi:2018oca}, and the limit on its decay branching ratio is given by
\be
{\rm BR}(B_d \to \mu^+\mu^- \gamma) < 1.6 \times 10^{-7}\,.
\label{limit3}
\ee
But there is no experimental limit on $B_s \to \mu^+\mu^- \gamma$ yet. Similar to the case of $B_d \to e^+e^- \gamma$, we notice that the upper limit on $B_d \to \mu^+\mu^- \gamma$ decay branching ratio is much larger than the SM prediction. 
The other experimental measurements relevant to the analysis are summarized in Table~\ref{T3}. 

\begin{table}[htb]
 \begin{center}
 \begin{tabular}{|c|c|c|c|} 
 \hline
 Decay mode & BaBar & Belle & LHCb \\ [0.5ex] 
 \hline
 $B^0 \to \pi^0 \mu^+\mu^-$ & $< 6.9 \times 10^{-8}$ & $< 1.8 \times 10^{-7}$ & - \\
 $B^+ \to \pi^+ \mu^+\mu^-$ & $< 5.5 \times 10^{-8}$ & $< 6.9 \times 10^{-8}$ & $(1.76 \pm 0.23) \times 10^{-8}$ \\ 
 \hline
 $B^0 \to K^0 \mu^+\mu^-$ & $4.9^{+2.9}_{-2.5} \pm 0.3$ & $4.4^{+1.3}_{-1.1} \pm 0.3$ & $3.27 \pm 0.34 \pm 0.17$ \\
 $B^+ \to K^+ \mu^+\mu^-$ & $4.1^{+1.6}_{-1.5} \pm 0.2$ & $5.3^{+0.8}_{-0.7} \pm 0.3$ & $4.29 \pm 0.07 \pm 0.21$ \\
 \hline
 $B^0 \to K^{*0} \mu^+\mu^-$ & $13.5^{+4.0}_{-3.7} \pm 1.0$ & $10.6^{+1.9}_{-1.4} \pm 0.7$ & $9.04^{+0.16}_{-0.15} \pm 0.62$ \\
 $B^+ \to K^{*+} \mu^+\mu^-$ & $14.6^{+7.9}_{-7.5} \pm 1.2$ & $11.1^{+3.2}_{-2.7} \pm 1.0$ & $9.24 \pm 0.93 \pm 0.67$ \\
 \hline
 $B_s \to \overline{K}^{*0} \mu^+\mu^-$ & - & - & $(2.9 \pm 1.1) \times 10^{-8}$ \\
 \hline
 \end{tabular}
\end{center}
\caption{Observed experimental results on $B \to P/V \mu^+\mu^-$ and $B_s \to \overline{K}^{*0} \mu^+\mu^-$ decay branching ratios from BaBar \cite{Aubert:2008ps,Lees:2013lvs}, Belle \cite{Wei:2009zv,Wei:2008nv} and LHCb \cite{Aaij:2014pli,Aaij:2015nea,Aaij:2016flj,Aaij:2018jhg}. The results for $B \to K^{(*)} \mu^+ \mu^-$ are shown in units of $10^{-7}$.}
\label{T3}
\end{table}

\begin{figure}[tb]
  \begin{center}
    \includegraphics[width=0.45\textwidth,bb= 0 0 420 427]{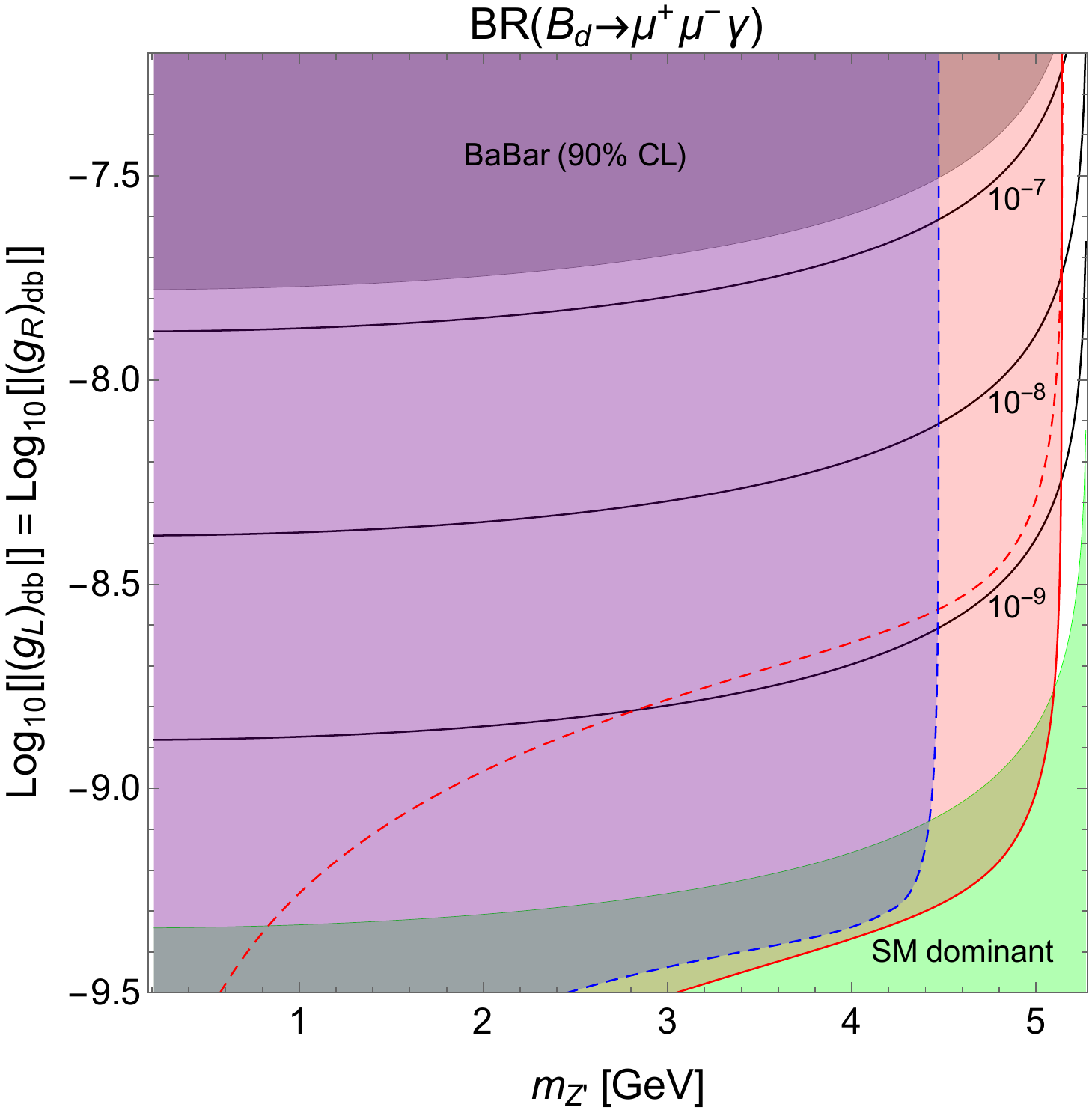} \hspace{1.0em}
    \includegraphics[width=0.45\textwidth,bb= 0 0 420 427]{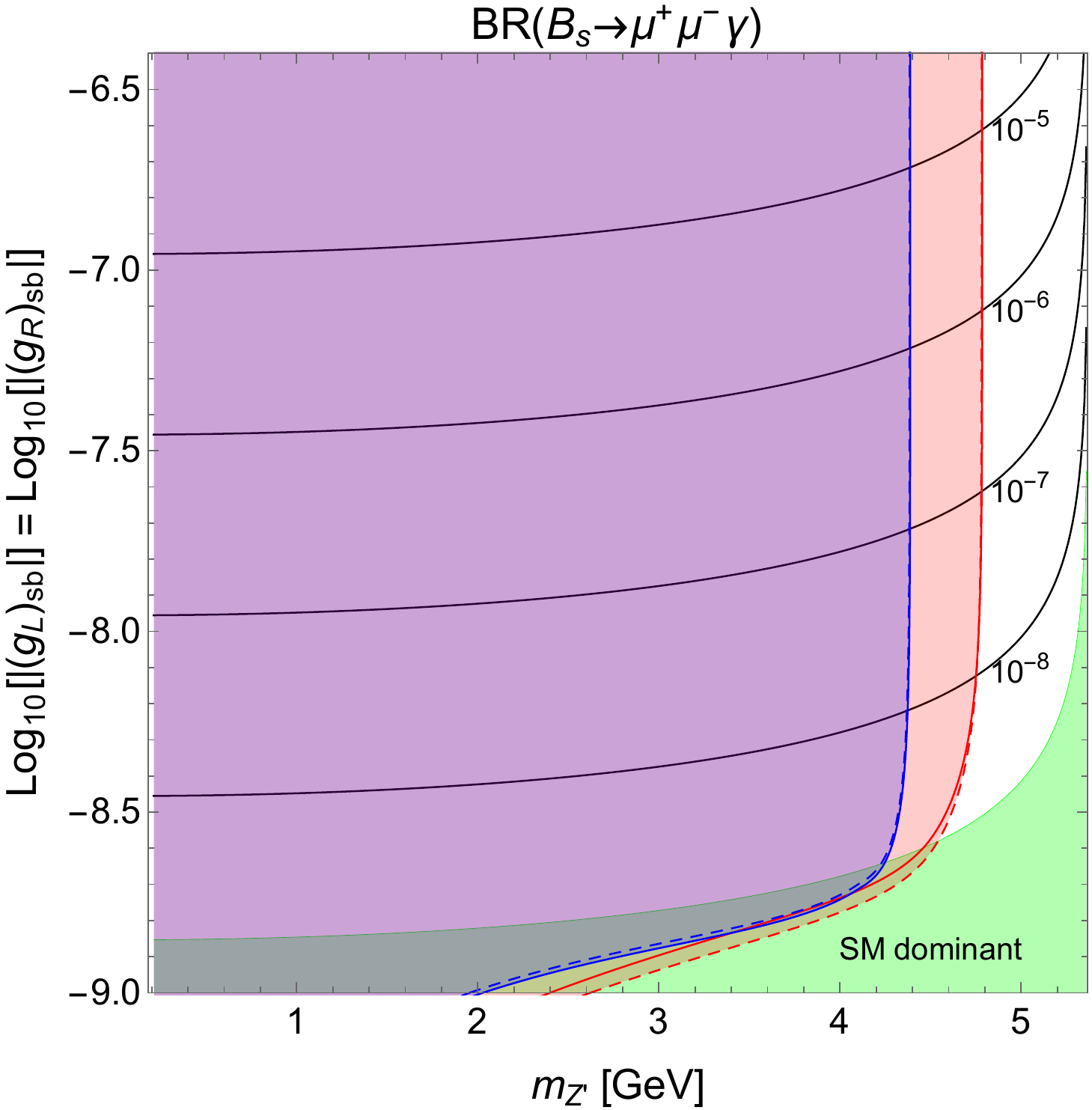}
    \caption{Branching ratios of $B_q \to \mu^+\mu^- \gamma$. Left panel is for $B_d \to \mu^+\mu^- \gamma$, and right panel is for $B_s \to \mu^+\mu^- \gamma$. The meanings of each color plot are the same as Fig.~\ref{fig:BtogamMET}.}
    \label{fig:Btogammumu}
  \end{center}
\end{figure}

We show the results in Fig.~\ref{fig:Btogammumu}. 
The left panel is for $B_d \to \mu^+\mu^- \gamma$, and the right panel for $B_s \to \mu^+\mu^- \gamma$. The meanings of each color plot are the same as Fig.~\ref{fig:BtogamMET}, and the same assumption about the ``rough bound" as $B_q \to e^+e^- \gamma$ is imposed if there is already an observed value for $B \to P \mu^+\mu^-$ or $B \to V \mu^+\mu^-$. 
In contrast to the electron case, the LHCb collaboration gives a strong constraint on $g_A^{(d)}$ from $B_s \to \overline{K}^{*0} \mu^+\mu^-$ \cite{Aaij:2018jhg} which is caused by $b \to d$ transition~\footnote{Recall that in Table~\ref{T2} we did not list the LHCb data of $B_s \to \overline K^{*0} e^+e^-$. The reason is attributed to the relatively low tagging efficiency of electron at the LHCb detector. Moreover, for any modes there are no data of $B_s$ decay from Belle, because its center-of-momentum energy is at the mass of the $\Upsilon(4S)$ resonance, whose decay into a pair of $B_s$ is kinematically suppressed.}. This bound is shown by the blue dashed line in the left panel. For $B_s \to\mu^+\mu^-\gamma$, both the vector and axial-vector couplings are strongly constrained by Belle, etc. Therefore, the couplings $(g_{L,R})_{qb}$ are severely constrained by $B \to P/V \mu^+\mu^-$, 
and we expect no new physics effects in $B_q \to \mu^+\mu^- \gamma$ decays.

\subsection{Explicit model case}
\subsubsection{Realistic branching ratios for $Z' \to \bar{f} f$}

\begin{figure}[tb]
  \begin{center}
    \includegraphics[width=0.75\textwidth,bb= 0 0 515 285]{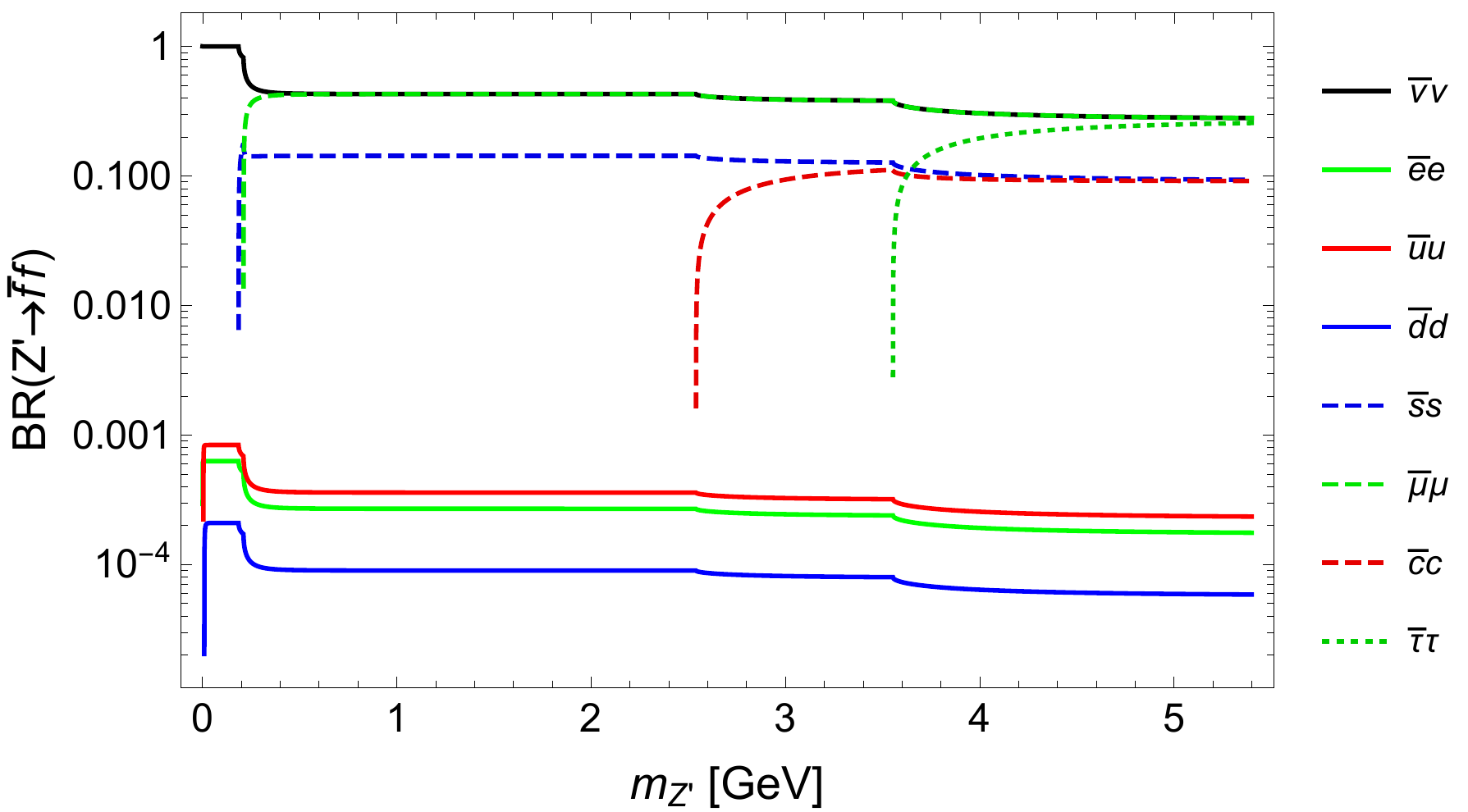}
    \caption{Branching ratios of $Z' \to \bar{f} f$ in the model taken from Ref.~\cite{Kang:2019vng}. Here, BR$(Z' \to \bar{\nu} \nu)$ means the sum of branching ratios for three neutrinos of the SM.}
    \label{fig:DecZp}
  \end{center}
\end{figure}

The above analysis is for the simplified model where $Z'$ decays only to one channel of lepton pair. 
However, if we consider an explicit model, $Z'$ decays in each channel with specific ratio defined by the model. 
Therefore, as an example, we will show the predictions for the model proposed in Ref.~\cite{Kang:2019vng}. 
We show each branching ratio for $Z' \to \bar{f} f$ in this model in Fig.~\ref{fig:DecZp}.
Note that for this figure, we assume that the couplings between $Z'$ and $\bar{q} q$ ($q = u, d, s, c$) are approximately vector-like ones, and flavor violating $Z'$ decays are neglected due to the smallness of its couplings. 
In addition, the couplings between $Z'$ and charged lepton pairs are also vector-like and depend only on one parameter, namely, gauge coupling $g'$. 
However, the branching ratios in Fig.~\ref{fig:DecZp} do not depend on $g'$ since all decay widths for $Z' \to \bar{f} f$ are proportional to ${g'}^2$. 
Moreover, the first generation of fermions has coupling to $Z'$ through a kinetic mixing term in this model.

\begin{figure}[tb]
  \begin{center}
    \includegraphics[width=0.6\textwidth,bb= 0 0 420 427]{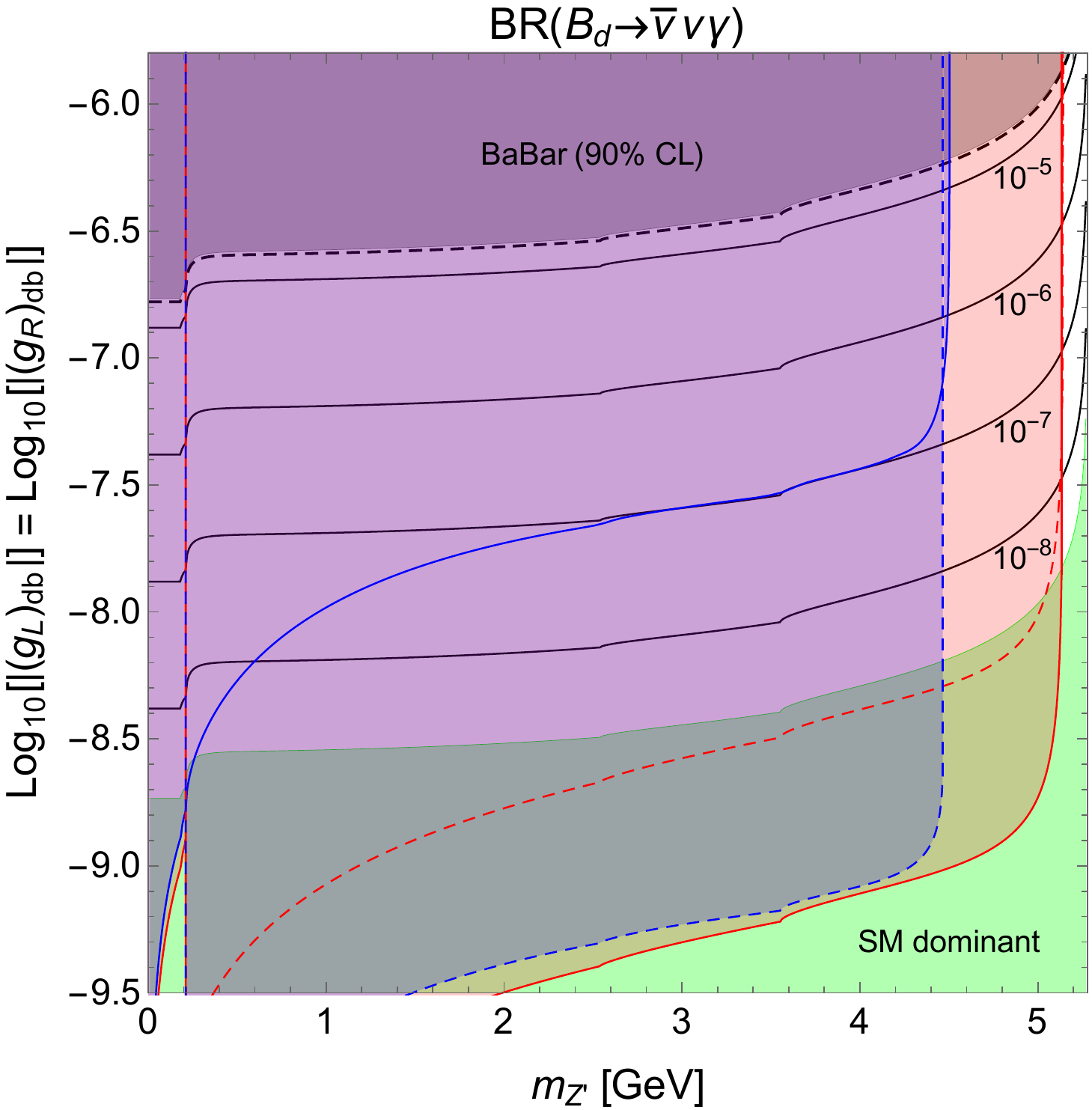}
    \caption{Branching ratios of $B_d \to \bar{\nu} \nu \gamma$ for the model in Ref.~\cite{Kang:2019vng}. The meanings of each color plot are the same as Fig.~\ref{fig:BtogamMET}, but all the constraints from $B \to P \bar{\ell} \ell$ and $B \to V \bar{\ell} \ell$ ($\ell = \nu, e, \mu$) are considered.}
    \label{fig:Bdtogamnunu}
  \end{center}
\end{figure}

In this case, we should take into account the all constraints from $B \to P \bar{\ell} \ell$ and $B \to V \bar{\ell} \ell$, shown in Tables~\ref{T1}, \ref{T2} and \ref{T3}, and hence, the flavor violating couplings are severely constrained by all the decay modes. 
From Fig.~\ref{fig:BtogamMET}, $B_d \to \bar{\nu} \nu \gamma$ has the possibility to be much larger than the SM prediction, as an example, we show the results for this decay mode in Fig.~\ref{fig:Bdtogamnunu}. 
The meanings of each color plot are the same as Fig.~\ref{fig:BtogamMET}, and we also assume the ``rough bound" for some decay modes, mentioned above. 
In this case, the strong constraints come from as follows: 
for $(g_L)_{db} = (g_R)_{db}$ case, $B^+ \to \pi^+ \bar{\nu} \nu$ ($m_{Z'} \leq 2 m_{\mu}$), $B^+ \to \pi^+ \mu^+\mu^-$ ($2 m_{\mu} < m_{Z'} < M_{B^+} - M_{\pi^+}$) and $B^0 \to \pi^0 \mu^+\mu^-$ ($M_{B^+} - M_{\pi^+} \leq m_{Z'} < M_{B^0} - M_{\pi^0}$); 
for $(g_L)_{db} = - (g_R)_{db}$ case, $B^+ \to \rho^+ \bar{\nu} \nu$ ($m_{Z'} \leq 2 m_{\mu}$ and $M_{B_s} - M_{K^{*0}} \leq m_{Z'} < M_{B^+} - M_{\rho^+}$) and $B_s \to \overline{K}^{*0} \mu^+\mu^-$ ($2 m_{\mu} < m_{Z'} < M_{B_s} - M_{K^{*0}}$). 

As expected, $(g_{L,R})_{db}$ are severely constrained. 
In particular, when $m_{Z'} > 2 m_{\mu}$, the decay modes involving muon pair give the strong constraint on the flavor violating couplings. 
Therefore, we cannot expect any specific signature in $B_d \to \bar{\nu} \nu \gamma$ decay mode unless we suppress $B \to P/V \mu^+ \mu^-$ decays, 
for example assuming tiny $Z'$ coupling to the muon in order to make BR($Z' \to \mu^+ \mu^-$) smaller. 
Note that the decay modes involving electron pair do not give strong constraints in this model since the $Z'$ coupling to the electron is suppressed by the kinetic mixing. 
If there is no such suppression in the coupling, one should take into account the constraints from related processes. 
Furthermore, because of this kinetic mixing, BR$(B_d \to e^+e^- \gamma)$ cannot be large, and almost all the area for $m_{Z'} \lesssim 5$ GeV is constrained by other $B$ meson decay processes.

\subsubsection{Lepton flavor violating $Z'$ couplings}

In the above analyses, we assume that there are no LFV couplings of $Z'$. 
However, these couplings may exist in general flavorful $Z'$ model, although these are severely constrained by experimental results for LFV processes, like $\ell_i \to \ell_j \gamma$. 
Actually, the lepton sector of the model in Ref.~\cite{Kang:2019vng} should be extended to accommodate realistic neutrino mixing angles. 
In such an extension, the LFV couplings will be induced, and we can obtain some specific prediction of LFV processes. 
Therefore, we also investigate the prediction of radiative $B$ meson decays with $Z' \to \ell_i^+ \ell_j^-$ ($\ell = e, \mu, \tau; i \neq j$). 
Note that the SM predictions for these decay modes are suppressed because of the absence of LFV in the SM, and at present, there are no direct bounds for such decay modes. 

First of all, let us consider the simplified case where $Z'$ only decays to $\ell_i^+ \ell_j^-$, namely, BR$(Z' \to \ell_i^+ \ell_j^-) = 1$. 
Similar to the above analyses, we should take into account the constraints from $B \to P \ell_i^{\pm} \ell_j^{\mp}$ and $B \to V \ell_i^{\pm} \ell_j^{\mp}$, which are summarized in Refs.~\cite{Tanabashi:2018oca,Amhis:2019ckw}. 
Roughly speaking, the upper bounds for $B \to P(V) e^{\pm} \mu^{\mp}$ are $\mathcal{O}(10^{-7})$, while those for $B \to P(V) \ell^{\pm} \tau^{\mp}$ ($\ell = e$ or $\mu$) are $\mathcal{O}(10^{-4} \mathchar`- 10^{-5})$. 
The latter bounds are similar to the bounds on $B \to P(V) \bar{\nu} \nu$ in Table~\ref{T1}, and therefore, the predictions for $B_q \to \gamma \ell^{\pm} \tau^{\mp}$ are similar to Fig.~\ref{fig:BtogamMET}, which means that BR($B_q \to \ell^{\pm} \tau^{\mp} \gamma$) is estimated to be $\mathcal{O}(10^{-7})$. 
On the other hand, BR($B_q \to e^{\pm} \mu^{\mp} \gamma$) is at least two orders of magnitude smaller than BR($B_q \to \ell^{\pm} \tau^{\mp} \gamma$) because of its bounds. 

However, when LFV couplings of $Z'$ exist, it is natural that lepton flavor conserving couplings $\ell_i \ell_i Z'$ also exist. 
In this case, the simplified case is no longer valid, and we should consider explicit branching ratio for $Z' \to \ell_i^+ \ell_j^-$, as in Fig.~\ref{fig:DecZp}. 
Since the explicit branching ratio for $Z' \to \ell_i^{\pm} \ell_j^{\mp}$ is small because of the severe LFV bounds, the predictions become smaller than above estimation. 
For example, BR$(Z' \to \ell_i^+ \ell_j^-) = \mathcal{O}(10^{-9} \mathchar`- 10^{-10})$ when LFV couplings is $\mathcal{O}(10^{-4})$. 
Therefore, we cannot expect any interesting predictions for this model in $B_q \to \ell_i^{\pm} \ell_j^{\mp} \gamma$ decays unless the future experiments reach the level of BR$(B_q \to \ell_i^{\pm} \ell_j^{\mp} \gamma) \sim \mathcal{O}(10^{-16} \mathchar`- 10^{-18})$.

\section{Conclusions and discussions}
\label{sec:conclusion}

In this paper, we investigate the signatures of flavor changing processes induced by a light $Z'$ in $B_q \to \gamma Z'$ decays. If $Z'$ has the flavor violating couplings in $b \to q$ ($q = d, s$) transitions, $B_q \to \gamma Z'$ can be induced at the tree-level. 
We give the specific expression for the decay width of $B_q \to \gamma Z'$ in Eq.~\eqref{eq:GamBqtogammaZp}, 
and we also consider the case with cascade decays of an on-shell $Z'$. 

However, such flavor violating couplings are also related to the other flavor processes, for example $B \to P \ell^+ \ell^-$ and $B \to V \ell^+ \ell^-$ where $P$ and $V$ stand for pseudoscalar and vector mesons, respectively. 
In the simplified model where $Z'$ decays only to one channel, BR$(B_d \to \gamma + \ET)$ can be as large as $\mathcal{O}(10^{-7})$ because of the weak bounds from the $B \to P/V + \ET$ decays. 
On the other hand, for the charged lepton case, BR$(B_q \to \ell^+ \ell^- \gamma)$ ($\ell = e, \mu$) is severely constrained and is generally expected to be lower than the experimental sensitivities. 
Interestingly, however, one can avoid such bounds by considering $(g_L)_{db} = - (g_R)_{db}$ in $B_d \to e^+ e^- \gamma$, and its branching ratio can be as large as $\mathcal{O}(10^{-7})$. For the muon case, we cannot obtain the large branching ratio compared to the SM prediction unless $m_{Z'} \sim 5$ GeV because of the constraints put by the rich experimental measurements for relevant $B$ meson decay processes.

We also show the predictions for an explicit model constructed in Ref.~\cite{Kang:2019vng}. 
In this case, the $Z'$ decays to all $\bar{f} f$ channels if it is kinematically allowed, and hence, we should take into account all indirect bounds shown in Tables~\ref{T1}, \ref{T2} and \ref{T3} by considering appropriate branching ratios predicted by the model. 
Since the bounds from $B \to P/V \mu^+ \mu^-$ give strong constraints when $m_{Z'} > 2 m_{\mu}$, we cannot expect specific signatures for this model in $B_q \to \bar{\nu} \nu \gamma$ decays. 
Therefore, we should consider the suppression for these processes in order to enhance the branching ratio for $B_q \to \gamma Z'$ decays. 
In addition, we discuss the prediction for the radiative $B$ meson decays with LFV, BR($B_q \to \ell_i^{\pm} \ell_j^{\mp} \gamma$). 
In a simplified model where $Z'$ couples only to $\ell_i \ell_j$, although it may not be a realistic model, its branching ratios can be $\mathcal{O}(10^{-7} \mathchar`- 10^{-9})$. 

In conclusion of our analyses, the branching ratio of radiative $B$ meson decays $(B\to \gamma Z^{\prime})$ can be large at a level of $\mathcal{O}(10^{-7})$ when $Z'$ decays to $\bar{\nu} \nu$ (or totally invisible decay), $e^+ e^-$ or $\ell^{\pm} \tau^{\mp}$, relatively larger than the expectation of the SM.
Therefore, we can expect that there is room to search for the signatures of new physics around $10^{-7}$ of decay branching ratio, and hence, we suggest the LHCb and Belle II collaborators search for these signatures. 
The current direct bound for $B_d \to \bar{\nu} \nu \gamma$ is $\mathcal{O}(10^{-5})$, while that for $B_d \to e^+ e^- \gamma$ is just above the $10^{-7}$. 
A method to search for $B_s \to \mu^+ \mu^- \gamma$ from $B_s \to \mu^+ \mu^-$ developed by 
Dettori {\it et. al} in Ref.~\cite{Dettori:2016zff} can be applied to obtain the sensitivity of $B \to e^+ e^- \gamma$
decay from $B\to e^+ e^-$ decay. This would further constrain the study of rare $B$ decay performed in the present 
work if recent $B \to e^+ e^-$ decay result from LHCb \cite{Aaij:2020nol} is taken into account. 
However, a detailed study to evaluate $B \to e^+ e^- \gamma$ decay sensitivity from $B \to e^+ e^-$
decay is beyond the scope of present work. 
Moreover, it is worth searching for the mode of $B \to \ell_i^{\pm} \ell_j^{\mp} \gamma$ since the LFV is a clear signal for the new physics beyond the SM.

\vspace*{0.3cm}
\noindent{\bf Acknowledgements}

This work is supported in part by the National Science Foundation of China (11775086, 11775093, 11422545, 11947235).

\appendix

\section{Decay width for the off-shell $Z'$ case}
The decay width of $B_q \to \ell^+ \ell^- \gamma$ with off-shell $Z'$ can be calculated and the result is given by
\begin{align}
\Gamma (B_q \to \ell^+ \ell^- \gamma) = &\, \frac{\alpha}{512 \pi^2} \frac{M_{B_q}}{27} \frac{f_{B_q}^2}{\lambda_{B_q}^2} \Bigl( | g_V^{(q)} |^2 + | g_A^{(q)} |^2 \Bigr) \\
&\times \int_0^{1 - 4 r_{\ell}} \! d x_{\gamma} \frac{v x_{\gamma}}{(1 - x_{\gamma} - r_{Z'})^2} \Bigl[ | g_V^{(\ell)} |^2 (1 - x_{\gamma} + 2 r_{\ell}) + | g_A^{(\ell)} |^2 (1 - x_{\gamma} - 4 r_{\ell}) \Bigr], \nonumber
\end{align}
where $g_{V, A}^{(\ell)} \equiv (g_L)_{\ell \ell} \pm (g_R)_{\ell \ell}$, $v \equiv \sqrt{1 - 4 r_{\ell} / (1 - x_{\gamma})}$, $x_{\gamma} \equiv 2 E_{\gamma} / M_{B_q}$ and $r_{\ell} \equiv m_{\ell}^2 / M_{B_q}^2$. This decay width formula can be applied to the decay $B_q \to \bar{\nu} \nu \gamma$ by replacing $r_{\ell} \to r_{\nu} = m_{\nu}^2 / M_{B_q}^2 \approx 0$ and $g_{V, A}^{(\ell)} \to (g_L)_{\nu \nu}$. 
This result is consistent with the result in Ref.~\cite{Dincer:2001hu}.

\end{document}